\documentclass[conference]{IEEEtran}
\IEEEoverridecommandlockouts
\usepackage{cite}
\usepackage{amsmath,amssymb,amsfonts}
\usepackage{algorithmic}
\usepackage{graphicx}
\usepackage{textcomp}
\usepackage{xcolor}
\usepackage[absolute]{textpos}

\usepackage{enumitem}
\usepackage{multirow}
\usepackage{lipsum}
\usepackage{listings}
\usepackage{subfigure}
\usepackage{graphicx}
\usepackage{pdfpages}
\usepackage{verbatim}
\usepackage{pifont}
\usepackage{etoolbox}
\usepackage{textcomp}
\usepackage{ulem}
\usepackage{xcolor}
\usepackage{hyperref}
\usepackage{url}
\usepackage{caption}
\usepackage{booktabs} % For nice tables
\usepackage{siunitx} % To align table numbers by unit

\usepackage[switch]{lineno} % default option is 'left'

\usepackage[ruled,norelsize]{algorithm2e}

\def\BibTeX{{\rm B\kern-.05em{\sc i\kern-.025em b}\kern-.08em
    T\kern-.1667em\lower.7ex\hbox{E}\kern-.125emX}}

\newcommand\blfootnote[1]{%
  \begingroup
  \renewcommand\thefootnote{}\footnote{#1}%
  \addtocounter{footnote}{-1}%
  \endgroup
}

\begin{document}

\title{Accelerating Encrypted Computing on Intel GPUs}
%\title{Accelerating Homomorphic Encryption Based on the CKKS Scheme on Intel GPUs}

\author{\IEEEauthorblockN{Yujia Zhai,\IEEEauthorrefmark{1}
Mohannad Ibrahim,\IEEEauthorrefmark{2}
Yiqin Qiu,\IEEEauthorrefmark{3}
Fabian Boemer,\IEEEauthorrefmark{3}
Zizhong Chen,\IEEEauthorrefmark{1}
Alexey Titov,\IEEEauthorrefmark{3}
Alexander Lyashevsky,\IEEEauthorrefmark{3}}
\IEEEauthorblockA{\IEEEauthorrefmark{1}University of California, Riverside, CA, USA}
\IEEEauthorblockA{\IEEEauthorrefmark{2}North Carolina State University, Raleigh, NC, USA}
\IEEEauthorblockA{\IEEEauthorrefmark{3}Intel Corporation, Santa Clara, CA, USA}
yzhai015@ucr.edu, mmibrah2@ncsu.edu, yiqin.qiu@intel.com, fabian.boemer@intel.com, \\chen@cs.ucr.edu, alexey.titov@intel.com, alexander.lyashevsky@intel.com
}

\maketitle

\thispagestyle{plain}
\pagestyle{plain}

%\linenumbers

\begin{abstract}
Homomorphic Encryption (HE) is an emerging encryption scheme that allows computations to be performed directly on encrypted messages. This property provides promising applications such as privacy-preserving deep learning and cloud computing. Prior works have been proposed to enable practical privacy-preserving applications with architectural-aware optimizations on CPUs, GPUs and FPGAs. However, there is no systematic optimization for the whole HE pipeline on Intel GPUs. In this paper, we present the first-ever SYCL-based GPU backend for Microsoft SEAL APIs. We perform optimizations from instruction level, algorithmic level and application level to accelerate our HE library based on the Cheon, Kim, Kim and Song (CKKS) scheme on Intel GPUs. The performance is validated on two latest Intel GPUs. Experimental results show that our staged optimizations together with optimizations including low-level optimizations and kernel fusion accelerate the Number Theoretic Transform (NTT), a key algorithm for HE, by up to 9.93X compared with the naive GPU baseline. The roofline analysis confirms that our optimized NTT reaches 79.8\% and 85.7\% of the peak performance on two GPU devices. Through the highly optimized NTT and the assembly-level optimization, we obtain 2.32X - 3.05X acceleration for HE evaluation routines. In addition, our all-together systematic optimizations improve the performance of encrypted element-wise polynomial matrix multiplication application by up to 3.10X.

\end{abstract}

\begin{IEEEkeywords}
Homomorphic Encryption, Number Theoretic Transform, Intel GPU, CKKS, Privacy-Preserving Computing\blfootnote{©Notice: Copyright ©2021, Intel Corporation. All Rights Reserved. Intel TM Notice:
Intel, the Intel logo, and other Intel marks are trademarks of Intel Corporation or
its subsidiaries. Other names and brands may be claimed as the property of others.
No product or component can be absolutely secure. Performance/Benchmarking Disclaimer: Performance varies by use, configuration and other factors. Learn more at
www.intel.com/performanceindex. Intel technologies may require enabled hardware,
software or service activation. Your results may vary. No product or component can
be absolutely secure. Performance results are based on testing as of dates shown in
configurations and may not reflect all publicly available updates.}
\end{IEEEkeywords}

\section{Introduction}

%Data privacy is the main barrier for enterprises to offload computing tasks to the cloud. 
The COVID-19 pandemic boosts the rapidly growing demand of enterprises on cloud computing. By 2021, 50\% of enterprise workloads are deployed to public clouds and this percentage is expected to breach 57\% in the next 12 months \cite{flexera}. Although outsourcing data processing to cloud resources enables enterprises to relieve the overhead of deployment and maintenance for their private servers, it raises security and privacy concerns of the potential sensitive data exposure.

Adopting traditional encryption schemes to address this privacy concern is less favorable because a traditional encryption scheme requires decrypting the data before the computation, which presents a vulnerability and may destroy the data privacy. In contrast, Homomorphic Encryption (HE), an emerging cryptographic encryption scheme, is considered to be one of the most promising solutions to such issues. HE allows computations to be performed directly on encrypted messages without the need for decryption. Therefore, this encryption scheme protects private data from both internal malicious employees and external intruders, while assuming honest computations.

In 1978, Rivest, Adleman, and Dertouzous \cite{rivest1978data}, first introduced the idea of computing on encrypted data through the use of ``privacy homomorphisms". Since then, several HE schemes have been invented, which can be categorized by the types of encrypted computation they support.
\textit{Partial} HE schemes enable only encrypted additions or multiplications. The famous RSA cryptosystem is, in fact, the first HE scheme, supporting encrypted modular multiplications.
% The past decades have witnessed the rapid improvement of theoretical studies on homomorphic encryption schemes.
% % For a pair of public keys \{$e,n$\}, \textbf{Enc}($m$)=$m^e \mod n$. Multiplying two encrypted messages maintains the homomorphism because \textbf{Enc}($m_1$)$\cdot$\textbf{Enc}($m_2$)$ = m^e_1\cdot m^e_2 \mod n = $ \textbf{Enc}($m_1\cdot m_2$).
% One can quickly verify that the addition for an RSA-encrypted ciphertext does not maintain this property.
% RSA is thus a \textit{partial} homomorphic encryption scheme, supporting only one encrypted operator.
In contrast, the Paillier cryptosystem \cite{paillier1999public} is a partial HE scheme that supports only modular additions.

\textit{Levelled} HE schemes, on the other hand, support both encrypted additions and multiplications, but only up to a certain circuit depth $L$ determined by the encryption parameters. The Brakerski/Fan-Vercauteren (BFV) \cite{fan2012somewhat} and Brakerski-Gentry-Vaikuntanathan (BGV) \cite{brakerski2014leveled} schemes are two popular leveled HE schemes used today, which support exact integer computation.
In \cite{cheon2017homomorphic}, Cheon, Kim, Kim and Song presented the CKKS scheme, which treats the encryption noise as part of approximation errors that occur during computations within floating-point numerical representation. This imprecision requires a refined security model \cite{li2021security}, but provides faster runtimes than BFV/BGV in practice.
%making it suitable for use cases which do not require exact arithmetic. CKKS also has a slightly different security model is also vulnerable to attacks that BFV/BGV are not, requiring careful evaluation before deployment \cite{li2021security}.

\textit{Fully} HE schemes enable an unlimited number of encrypted operations, typically by adding an expensive bootstrapping step to a levelled HE scheme, as first detailed by Craig Gentry \cite{gentry2009fullyThesis}. TFHE \cite{chillotti2016tfhe} improves the runtime of bootstrapping, but requires evaluating circuits on binary gates, which becomes expensive for standard 32-bit or 64-bit arithmetic.

The improved capabilities and performance of these HE schemes have enabled a host of increasingly sophisticated real-world privacy-preserving applications. Early applications included basic statistics and logistic regression evaluation \cite{naehrig2011can}. More recently, HE applications have expanded to a wide variety of applications, including privatized medical data analytics, privacy-preserving machine learning, homomorphic dynamic programming, private information retrieval, and sorting on encrypted data\cite{bos2014private, cheon1439secure, boemer2019ngraph, rathee2020cryptflow2, graepel2012ml, angel2018pir, conglabeled2021, cetin2021homomorphic, chatterjee2013accelerating}.

The memory and runtime overhead of HE is a major obstacle to immediate real-world deployments.
Several HE libraries support efficient implementations of multiple HE schemes, including Microsoft SEAL \cite{laine2017simple} (BFV/CKKS), HElib \cite{halevi2013design} (BFV/BGV/CKKS), and PALISADE \cite{polyakov2017palisade} (BGV/BFV/CKKS/TFHE). In \cite{IntelHEXL}, Intel published HEXL, accelerating HE integer arithmetic on finite fields by featuring Intel Advanced Vector Extensions 512 (Intel AVX512) instructions. Since GPUs deliver higher memory bandwidth and computing throughput with lower unit power consumption, researchers presented libraries such as cuHE \cite{dai2015cuhe}, TFHE \cite{chillotti2016tfhe} and NuFHE \cite{nufhe} to accelerate HE using CUDA-enabled GPUs. Since GPUs deliver higher memory bandwidth and computing throughput with lower unit power consumption, researchers have also accelerated HE using CUDA-enabled GPUs, notably in the cuHE \cite{dai2015cuhe} and NuFHE \cite{nufhe} libraries.

Although HE optimizations on CPUs and CUDA-enabled GPUs have been extensively studied, an architecture-aware HE study investigating Intel GPUs has not been available. In addition, since the Number Theoretic Transform (NTT) is a crucial algorithm enabling fast HE computations, previous works on accelerating HE libraries majorly focus on optimizing computing kernels of NTT and inverse NTT (iNTT). However, a HE library supports basic HE primitives such encoding, encryption and evaluation. Such a comprehensive pipeline requires systematic optimizations, rather than only optimizing for computing kernels. In this paper, we present a HE library optimized for Intel GPUs based on the CKKS scheme. We not only provide a set of highly optimized computing kernels such as NTT and iNTT, but also optimize the whole HE evaluation pipeline at both the instruction level and application level. More specifically, our contributions include:

\begin{itemize}
    \item To the best of our knowledge, we design and develop the first-ever SYCL-based GPU backend for Microsoft SEAL APIs, which is also the first HE library based on the CKKS scheme optimized for Intel GPUs.
    \item We provide a staged implementation of NTT leveraging shared local memory of Intel GPUs. We also optimize NTT by employing strategies including high-radix algorithm, kernel fusion, and explicit multiple-tile submission.
    \item From the instruction level, we enable low-level optimizations for 64-bit integer modular addition and modular multiplication using inline assembly. We also provide a fused modular multiplication-addition operation to reduce the number of costly modular operations.
    \item From the application level, we introduce the memory cache mechanism to recycle freed memory buffers on device to avoid the run-time memory allocation overhead. We also design fully asynchronous HE operators and asynchronous end-to-end HE evaluation pipelines.
    \item We benchmark our HE library on two latest Intel GPUs. Experimental results show that our NTT implementations reaches up to 79.8\% and 85.7\% of the theoretical peak performance on both experimental GPUs, faster than the naive GPU baseline by 9.93X and 7.02X, respectively.
    \item Our NTT and assembly-level optimizations accelerate five HE evaluation routines under the CKKS scheme by 2.32X - 3.06X. In addition, the polynomial element-wise matrix multiplication applications are accelerated by 2.68X - 3.11X by our all-together systematic optimizations.
\end{itemize}

 The rest of the paper is organized as follows: we introduce background and related works in Section \ref{sec:background}, and then detail the asynchronous design and systematic optimization approaches in Section \ref{sec:design_and_optimizations}. Evaluation results are given in Section \ref{sec:results}. We conclude our paper and present future work in Section \ref{sec:conclusion}.

\section{Background and Related Works} \label{sec:background}

In this section, we briefly introduce the basics of the CKKS HE scheme. We then introduce the general architecture of Intel GPUs. Meanwhile, we summarize prior works of NTT optimizations on both CPUs and GPUs.

\subsection{Basics of CKKS} \label{sec:ckks}
% Follows presentation in https://eprint.iacr.org/2016/421.pdf

The CKKS scheme was first introduced in \cite{cheon2017homomorphic}, enabling approximation computation on complex numbers. This approximate computation is particularly suitable for real-world floating-point data, which is already inexact. Further work improved CKKS to support a full residue number system (RNS) \cite{cheon2018full} and bootstrapping \cite{cheon2018bootstrapping}. In this paper, we select CKKS as our FHE scheme, as implemented in Microsoft SEAL \cite{laine2017simple}.

Many HE schemes, including CKKS, BFV, and BGV, derive their security from the hardness of the ring learning with errors (RLWE) problem, which relates polynomials in a quotient ring. Let $\Phi_M[x]$ be the $M$'th cyclotomic polynomial, with \textit{polynomial modulus degree} $N = \phi(M).$ Then, given a \textit{ciphertext modulus} $q$, the ring $\mathcal{R}_q = \mathbb{Z}_q[x] / \Phi_M[x]$  consists of degree $N-1$ polynomials with integer coefficients in $\mathbb{Z}_q = \{0,1, \hdots, q-1\}$. For efficient implementation, $M$ is typically chosen to be a power of two, in which case $\Phi_M[x] = x^{M/2}+1 = x^N+1$ with $N$ also a power of two.
% $\mathbb{Z}_q[x]/(x^N+1)$, represents a set of polynomials modulus $q$ dividing over $x^N+1$.

CKKS has the message space consisting of $N/2$-dimensional complex vectors $\mathbb{C}^{N/2}$, plaintext space $\mathcal{R}_q$, and ciphertext space $\mathcal{R}_q^2$.\footnote{More generally, the ciphertext space can expand to more than two polynomials, but we restrict it to two polynomials to simplify our presentation.} We note multiplication in $\mathcal{R_q}$ includes a reduction by $X^N+1$ which ensures the products is also a degree $N-1$ polynomial. CKKS also takes in a precision argument $\Delta$ used to scale inputs. The ciphertext modulus is chosen as $q = q_L = \Delta^L\cdot q_0$. We denote $q_l = \Delta^l q_0$.

The CKKS scheme is composed of following basic primitives: \textbf{KeyGen}, \textbf{Encode}, \textbf{Decode}, \textbf{Encrypt}, \textbf{Decrypt}, \textbf{Add}, \textbf{Multiply} (\textbf{Mul}), \textbf{Relinearize} (\textbf{Relin}) and \textbf{Rescale} (\textbf{RS}). We assume that the reader is already familiar with these logical blocks, and therefore we provide below only cursory descriptions, while deeper descriptions can be found elsewhere \cite{cheon2017homomorphic}.

\textbf{KeyGen}($\lambda$)$\rightarrow$\{$pk \in \mathcal{R}_q^2$,$sk \in \mathcal{R}_q$,$evk \in \mathcal{R}_{P \cdot q}^2$\} takes a security parameter as an input to generate a set of keys for CKKS scheme, namely the secret key ($sk$), public key ($pk$), and evaluation key ($evk$). $pk$ and $sk$ are used for encryption and decryption, respectively, while $evk$ is used for homomorphic evaluation operations, which are accelerated on Intel GPUs in this paper and will be elaborated on later.

\textbf{Encode}($z \in \mathbb{C}^{N/2},\Delta)\rightarrow m \in \mathcal{R}_q$ encodes an input $z$ to a message $m$. Since the encoding scheme might destroy some significant numbers, encoding includes a multiplication by $\Delta$ to keep a precision of $\frac{1}{\Delta}$.

\textbf{Decode}($m \in \mathcal{R}_{q_l}, \Delta) \rightarrow z \in \mathbb{C}^{N/2}$
is the inverse operation of \textbf{Encode}. It includes a division by the scaling factor $\Delta$.

\textbf{Encrypt}($m \in \mathcal{R}_{q_L} ,pk$)$\rightarrow c \in \mathcal{R}_{q_L}^2$ takes an input plaintext $m$ and a public key $pk$. By sampling a vector $v$ and two Gaussian-distributed vectors $e_0,e_1$, the encrypted message is generated as $c=(v\cdot pk+(m+e_0,e_1)) \mod q_L$.

\textbf{Decrypt}($c \in \mathcal{R}_{q_l}^2,sk$)$\rightarrow m \in \mathcal{R}_q$ requires access to the secret key $sk$. The ciphertext $c = (c[0], c[1]) \in \mathcal{R}_{q_l}^2$ is decrypted as: $m'=c[0]+c[1]\cdot sk \mod q_l$.

\textbf{Add}($c_0 \in \mathcal{R}_{q_l}^2, c_1 \in \mathcal{R}_{q_l}^2$)$\rightarrow c_3 \in \mathcal{R}_{q_l}^2$ adds two input ciphertexts element-wise: $c_2 = (c_0[0] + c_1[0], c_0[1] + c_1[1]) \mod q_l$.

\textbf{Mul}($c_0 \in \mathcal{R}_{q_l}^2, c_1 \in \mathcal{R}_{q_l}^2$)$\rightarrow c_3 \in \mathcal{R}_{q_l}^2$ multiplies two input ciphertexts $c_0$ and $c_1$ to yield $c_2 = (c_0[0] c_1[0], c_0[0]c_1[1] + c_0[1]c_1[0], c_0[1]c_1[1]) \mod q_l$
%. For example, for both input ciphertexts of length 2 ($c_1=\{a_1,a_2\}$, $c_2=\{b_1,b_2\}$), the output ciphertext is equal to $c_3=\{a_1b_1,a_1b_2+a_2b_1,a_2b_2\}$, which is a ciphertext of length 3. After an ordinary polynomial multiplication, ciphertext multiplication requires to divide the result over $(x^N+1)$.

\textbf{Relin}($c \in \mathcal{R}_{q_l}^3,evk$)$\rightarrow c' \in \mathcal{R}_{q_l}^2$ decreases the size of the ciphertext back to (at least) 2 after multiplication. We have $c' = \left((c_0, c_1) + \left \lfloor P^{-1} \cdot c[2] \cdot \textit{evk} \right \rceil \right) \mod q_l$, where $P$ is a large auxiliary number used for basis conversion.

% We have \{$a'_1,a'_2\} = (a_1,a_2), p^{-1}\cdot a_2 \cdot evk$, where $a_2$ <<IS IT $a_2$ or $a_3$?<< is divided by a large integer $p$, rounded to the nearest integer and multiplied with the evaluation key $evk$. With similar procedures applied to $(a_1,a_2)$, <<IT's JUST APPLIED, WHAT DO YOU MEAN HERE?<< the length of original ciphertext is relinearized from $3$ to $2$.

% this was due to a typo. Thanks for pointing it out.

% \textbf{Relin}($\{a_1,a_2,a_3\},evk$)$\rightarrow \{a'_1,a'_2\}$ \textbf{Relin} decreases the size of the ciphertext back to (at least) 2 after multiplication. We have \{$a'_1,a'_2\} = \{(a_1,a_2) + \lfloor p^{-1}\cdot a_3 \cdot evk \rceil\}$, where $a_3$ is divided by a large integer $p$, rounded to the nearest integer and multiplied with the evaluation key $evk$.

%\textbf{RS} <<NOT CLEAR<< essentially applies a dividing-and-round operation to a ciphertext. Assuming the current scheme requires a relinearization after $L$ multiplications in maximal such that the noise does not exceed the bound, we have a modulus $q_L$ and all ciphertexts $c_i\in \mathbb{Z}_{q_L}$. After finishing one \textbf{Mul}, there remain $L-1$ permitted multiplications. Therefore, a new modulus $q_{L-1}$ can be computed and we need to rescale the ciphertext as \textbf{RS}$_{L\rightarrow L-1}=\left \lfloorTFHU \frac{q_{L-1}}{q_{L}} c \right \rfloor$ mod $q_{L-1}$.

\textbf{RS}${}_{l \rightarrow l'}$($c \in \mathcal{R}_{q_l}^2$)$\rightarrow c' \in \mathcal{R}_{q_{l'}}^2$ maintains the scale constant and reduces the noise in a ciphertext after \textbf{Mul}. However, it reduces the \textit{level} $l$ of the ciphertext, limiting the number of further homomorphic operations that can be performed.
\textbf{RS} computes $c' = \left \lfloor \frac{q_{l'}}{q_l} c \right \rceil \mod q_{l'}$
% Assuming the current scheme allows $L$ \textbf{Mul} operations in maximal such that the noise does not exceed the bound, we define the ciphertext modulus $q = \Delta^L\cdot q_0$.
Often, $l' = l-1$, in which case \textbf{RS}$_{l\rightarrow l-1}=\left \lfloor \Delta^{-1} c \right \rceil \mod q_{l-1}$. 

% As continuously performing \textbf{Mul}, the number of remaining permitted multiplications keeps decreasing until it equals zero. At a certain level $l$, we have modulus $q_l = \Delta^l\cdot q_0$. Then the rescaling operation from level $l$ to $l-1$ is defined as \textbf{RS}$_{l\rightarrow l-1}=\left \lfloor \frac{q_{l-1}}{q_{l}} c \right \rceil$ mod $q_{l-1}$, where $\lfloor \cdot \rceil$ indicates rounding to the nearest integer. Since $\left \lfloor \frac{q_{l-1}}{q_{l}} \right \rceil = \Delta^{-1}$, we have \textbf{RS}$_{l\rightarrow l-1}=\left \lfloor \Delta^{-1} c \right \rceil $ mod $q_{l-1}$.

\subsection{Number Theoretic Transform and Residue Number System} As noted in \cite{longa2016speeding}, the NTT can be exploited to accelerate multiplications in the polynomial ring $\mathcal{R}_q=\mathbb{Z}_q[x]/(x^N+1)$. We represent polynomials using a coefficient embedding: $\mathbf{a}=(a_0,...,a_{N-1})\in \mathbb{Z}_q^N$ and $\mathbf{b}=(b_0,...,b_{N-1})\in \mathbb{Z}_q^N$. Let $\omega$ be a primitive $N$-th root of unity in $\mathbb{Z}_q$ such that $\omega^N\equiv 1(\mod q)$. In addition, let $\psi$ be the 2$N$-th root of unity in $\mathbb{Z}_q$ such that $\psi^2=\omega$. Further defining $\Tilde{\mathbf{a}}=(a_0,\psi a_1,...,\psi^{N-1}a_{N-1})$ and $\Tilde{\mathbf{b}}=(b_0,\psi b_1,...,\psi^{N-1}b_{N-1})$, one can quickly verify that for $\mathbf{c}=\mathbf{a}\cdot \mathbf{b}\in \mathbb{Z}_q^N$, there holds the relationship $\mathbf{c}=\mathbf{\Psi^{-1}}\odot$ iNTT$($NTT$(\Tilde{\mathbf{a}})\odot$NTT $(\Tilde{\mathbf{b}}))$. Here $\odot$ denotes element-wise multiplication and $\mathbf{\Psi^{-1}}$ represents the vector $(1,\psi^{-1},\psi^{-2},...,\psi^{-(N-1)})$. Therefore, the total computational complexity of ciphertext multiplication in $\mathcal{R}_q$ is reduced from $O(N^2)$ to $O(N\log N)$.

In practice, since polynomial coefficients in the ring space are big integers under modulus $q$, multiplying these coefficients becomes computationally expensive. The Chinese Remainder Theorem (CRT) is typically employed to reduce this cost by transforming large integers to the Residue Number System (RNS) representation. According to CRT, one can represent the large integer $x \mod q$ using its remainders $(x \mod p_1, x \mod p_2, \hdots, x \mod p_n)$, where the moduli $(p_1,p_2,...,p_n)$ are co-prime such that $\Pi p_i=q$. We note the CKKS scheme has been improved from the initial presentation in Section \ref{sec:ckks} to take full advantage of the RNS \cite{cheon2018full}. 

To summarize what we have discussed, to multiply polynomials $\mathbf{a}$ and $\mathbf{b}$ represented as vectors in $\mathbb{Z}_q^N$, one needs to first perform the NTT to transform the negative wrapped $\Tilde{\mathbf{a}}$ and $\Tilde{\mathbf{b}}$ to the NTT domain. After finishing element-wise polynomial multiplication in the NTT domain, the iNTT is applied to convert the product to the coefficient embedding domain. When the polynomials are in RNS form, both the NTT and iNTT are decomposed to $n$ concurrent subtasks. Finally, we compute the outer product result by merging the iNTT-converted polynomial with $\mathbf{\Psi^{-1}}$.

\subsection{NTT optimizations} \label{sec:ntt}
Due to the pervasive usage of NTT and iNTT in HE, prior researchers proposed optimized implementations for NTT on CPUs\cite{IntelHEXL}, GPUs\cite{al2018high, goey2021accelerating, cuFHE, kim2020accelerating} and FPGAs\cite{kim2020hardware, riazi2020heax}. To be more specific, Intel HEXL provides a serial CPU implementation of the radix-2 negacyclic NTT using Intel AVX512 instructions\cite{IntelHEXL} and Harvey's lazy modular reduction approach \cite{harvey2014faster}. GPU-accelerated NTT implementations typically adopt the hierarchical algorithm first presented by Microsoft Research for the Discrete Fourier Transform (DFT)\cite{govindaraju2008high}. Among which, \cite{cuFHE} implements the hierarchical NTT with twiddle factors cached in shared memory. Rather than caching twiddle factors, \cite{kim2020accelerating} computes some twiddle factors on-the-fly to reduce the cost of modular multiplication and the memory access number of NTT. \cite{goey2021accelerating} further considers the built-in warp shuffling mechanism of CUDA-enabled GPUs to optimize NTT.

The hierarchical NTT implementation computes the NTT in three or four phases \cite{goey2021accelerating, govindaraju2008high}. An $N$-point NTT sequence is first partitioned into two dimensions $N = N_\alpha \cdot N_\beta$ and then $N_\alpha$ NTT workloads are proceeded simultaneously, where each workload computes an $N_\beta$-point NTT. After this column-wise NTT phase is completed, all elements are multiplied by their corresponding twiddle factors and stored to the global memory. In the next phase, $N_\beta$ simultaneous row-wise $N_\alpha$-point NTTs are computed followed by a transpose before storing back to the global memory. $N_{\alpha}$ and $N_{\beta}$ are selected to fit the size of shared memory on GPUs. Considering both the RNS representation of NTT and the batched processing opportunities in real-world applications can provide us with sufficient parallelisms, we adopt the staged NTT implementation rather than the hierarchical NTT implementation in this paper.

%Multiplying $\Tilde{a}\cdot \Tilde{b}$, we have $ c=\sum_k^Nc_kx^k$, where $c_k = \sum_{i=0}^ka_ib_{k-i}-\sum_{i=k+1}^{N-1}a_ib_{N+K-1}$. This expression to compute $c_k$ is called negacyclic convolution.

%The input polynomials, regardless ciphertexts or plaintexts, are transformed using NTT and then we perform a dyadic product between transformed polynomials. Of course we need to transform the computing results back through an inverse NTT after the multiplication is completed. 

%\subsubsection{Residue Number System (RNS)-enabled BFV scheme} As demonstrated in \cite{bajard2016full}, not only simply ring operations, division and round operations can also be performed by leveraging the Chinese Remainder Theorem (CRT), which suggests all aforementioned operations can be computed in a factorized modulus manner separately.

 %Mathematically, to compute $c \leftarrow a $ mod $p$, where $0\leq a\leq p$, one need to compute through $c \leftarrow a - \left \lfloor \frac{a}{p} \right \rfloor$. This intuitive method requires a division operation, which costs at least hundreds of clock cycles on modern GPUs \cite{arafa2019low}. To avoid the expensive division operations, one can adopt an algorithm called Barret reduction, which approximates the quotient $\left \lfloor \frac{a}{p} \right \rfloor$ using a precomputed $\frac{m}{2^k}\approx \frac{1}{p}$. In practical, we adopt $m = \left \lfloor \frac{2^k}{p} \right \rfloor$

\subsection{An Overview of Intel GPUs}

We use the Intel Gen11 GPU as an example to present the architecture Intel GPUs \cite{gen11arch}. Intel GPUs consist of a hierarchical architecture of computing and storage units. Starting from the low level, an Intel GPU contains a set of execution units (EU), where each EU supports up to seven simultaneous hardware threads, namely EU threads. In each EU, there is a pair of 128-bit SIMD ALUs, which support both floating-point and integer computations. Each of these simultaneous hardware threads has a 4KB general register file (GRF). So, an EU contains $7 \times 4$KB $=28 $KB GRF. Meanwhile, GRF can be viewed as a continuous storage area holding a vector of 16-bit or 32-bit floating-point or integer elements. For most Intel Gen11 GPUs, 8 EUs are aggregated into 1 Subslice. EUs in each Subslice can share data and communicate with each other through a 64KB highly banked data structure --- shared local memory (SLM). SLM is accessible to all EUs in a Subslice but is private to EUs outside of this Subslice. Not only supporting a shared storage unit, Subslices also possess their own thread dispatchers and instruction caches. Eight Subslices further group into a Slice, while additional logic such as geometry and L3 cache are integrated accordingly.

\section{Designs and Optimizations} \label{sec:design_and_optimizations}

\begin{figure}[ht]
\centering
\includegraphics[width=0.46\textwidth]{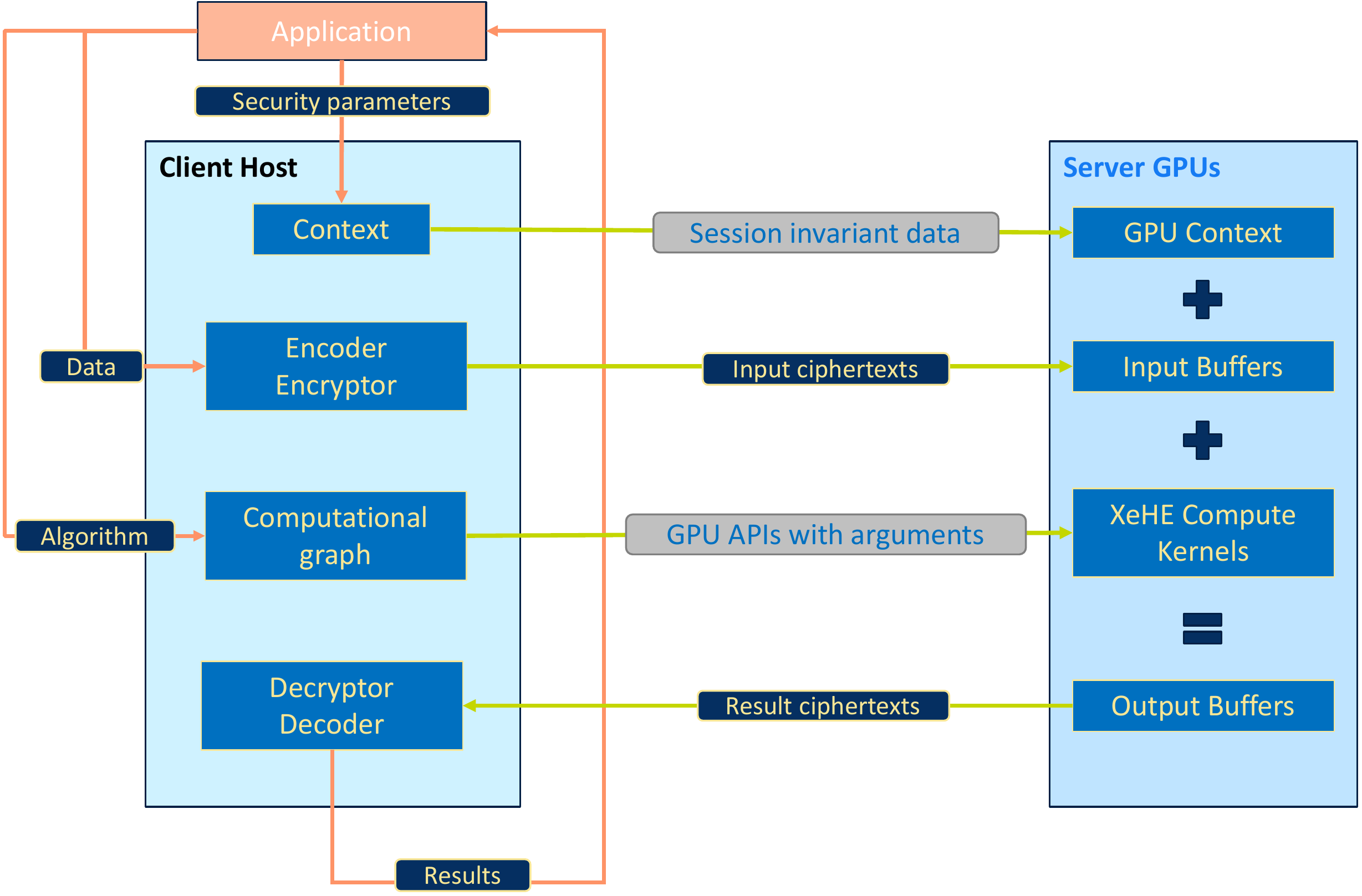}
\caption{Client(host)/Server (gpu) control/data flow.}
\label{fig:gpu-async-cfg}
\end{figure}

In the CKKS scheme, an input message is first encoded and then encrypted to generate ciphertexts using the public key provided by the key generation primitive. We evaluate (compute) directly on the encrypted messages (ciphertexts). Once the all computations are completed, the results can be decrypted and decode by the private key's owner. Figure \ref{fig:gpu-async-cfg} describes the control flow of our asynchronous HE library. Our HE library accelerates the HE evaluation using Intel GPUs while leaving other phases such as key generation, encoding, encryption, decryption and decoding on CPU. 

\begin{figure}[ht]
\centering
\includegraphics[width=0.46\textwidth]{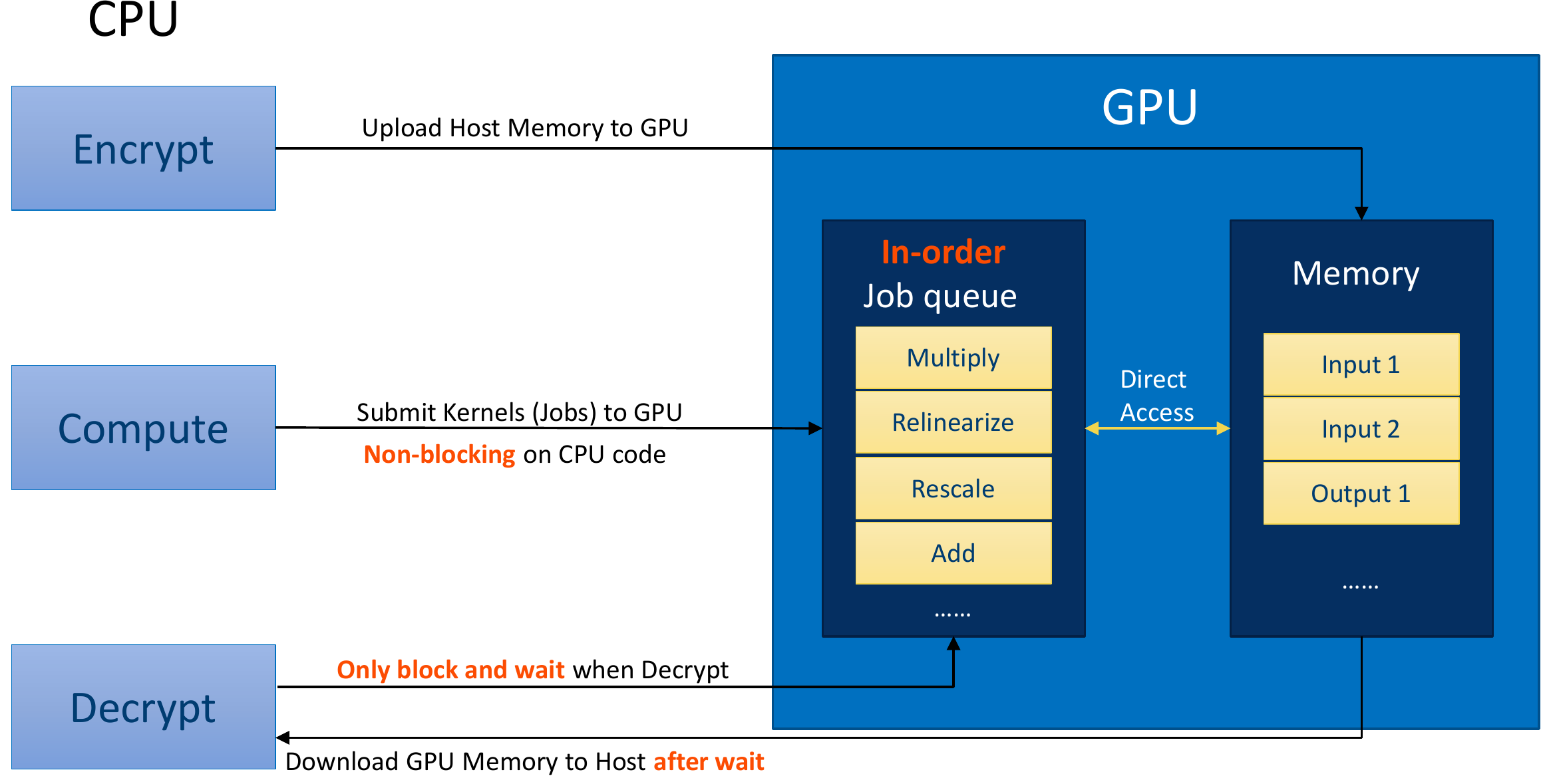}
\caption{Asynchronous execution scheme}
\label{fig:gpu-async}
\end{figure}

Once all the inputs and static data are sent to the GPU, the synchronization with the host becomes unnecessary. Since all host-device synchronizations take additional time, we developed a fully asynchronous execution pipeline to economize on synchronizations. As shown in Figure \ref{fig:gpu-async}, the computation on the GPU starts as soon as the first kernel of the computational graph is submitted. Meanwhile, GPU buffers are allocated and managed at runtime. GPU synchronizes with the host only after the buffers with the results are transferred back to the system memory. In the following contents of this section, we present optimizations of our library from three different angles: instruction, algorithm and application. 

\subsection{Instruction-level Optimizations}

Our HE library supports basic instructions such as addition, subtraction, multiplication and modular reduction -- all are 64-bit integer (int64) operations. We explicitly select int64 because our goal has been to provide accelerated SEAL APIs on Intel GPUs almost transparently. This is the reason why our current top-level software does not exactly fit to drive 32-bit integer (int32) calculations, although we envision to support both int32 and int64 eventually. Among these operations, the most expensive are modulus-related operations such as modular addition and modular multiplication. Although we can accelerate modular reduction using the Barrett reduction algorithm, which transforms the division operation to the less expensive multiplication operation, modular computations remain costly since no modern GPUs support int64 multiplication natively. Such multiplications are emulated at software level with the compiler support.

Based on these observations, we propose instruction-level optimizations from two aspects: 1) fusing modular multiplication with modular addition to reduce the number of modulo operations and 2) optimizing modular addition/multiplication from assembly level to remedy the compiler deficiency.

\subsubsection{Fused modular multiplication-addition operation}
 %Rather than eagerly applying modulus operation after both multiplication and addition, we propose to perform only one modulus operation after a \textit{pair} of consecutive multiplication and addition operations, namely a \textit{mad\_mod} operation. The overflow issue is not a concern when the both operands of multiplication are integers strictly less than 64 bits. <<This assumption holds because to assure a faster NTT transform, discussed in the next section, all of our ciphertexts are in the <<complex<< ring space under a integer modulus not exceeding 61 bits, that's all ciphertext elements do not exceed 61 bit as well<<.

 Rather than eagerly applying modulo operation after both multiplication and addition, we propose to perform only one modulo operation after a \textit{pair} of consecutive multiplication and addition operations, namely a \textit{mad\_mod} operation.We store the output of int64 multiplication in an 128-bit array. The potential overflow issue introduced by cancelling a modulus after addition is not a concern when both operands of addition are integers strictly less than 64 bits. This assumption holds because to assure a faster NTT transform, using David Harvey's optimizations \cite{harvey2014faster} following SEAL and discussed in the next section, all of our ciphertexts are in the ring space under a integer modulus less than 60 bits. Therefore, all ciphertext elements do not exceed 60 bits.

\subsubsection{Optimizing HE arithmetic operations using inline assembly}

%<<Inline assembly is a feature available in some compilers allowing programmers to embed low-level assembly code into their program written in higher-level languages such as C/C++<<.

Inline assembly is a feature available in some compilers allowing programmers to embed low-level assembly code into their program written in higher-level languages such as C/C++. Software engineers mainly resort to using inline assembly to optimize performance-sensitive parts of their programs. This is achieved through embedding a more efficient assembly implementation compared to what the compiler auto-generates. Additionally, it gives access to processor-specific or special instructions used to implement locking, synchronization and other specialized features that are not yet supported by the compiler.

The first step the process is to study and carefully review how Intel's DPC++ compiler generates code for core operations. Investigating the compiler’s generated code is key to low-level optimizations and could provide many opportunities for speedups and optimizations. We saw such opportunities in two of our core arithmetic operations: Unsigned Modular Addition, and Unsigned Integer Multiplication.

\paragraph{Unsigned Modular Addition (add\_mod)}

\begin{figure}[ht]
\centering
\includegraphics[width=0.4\textwidth]{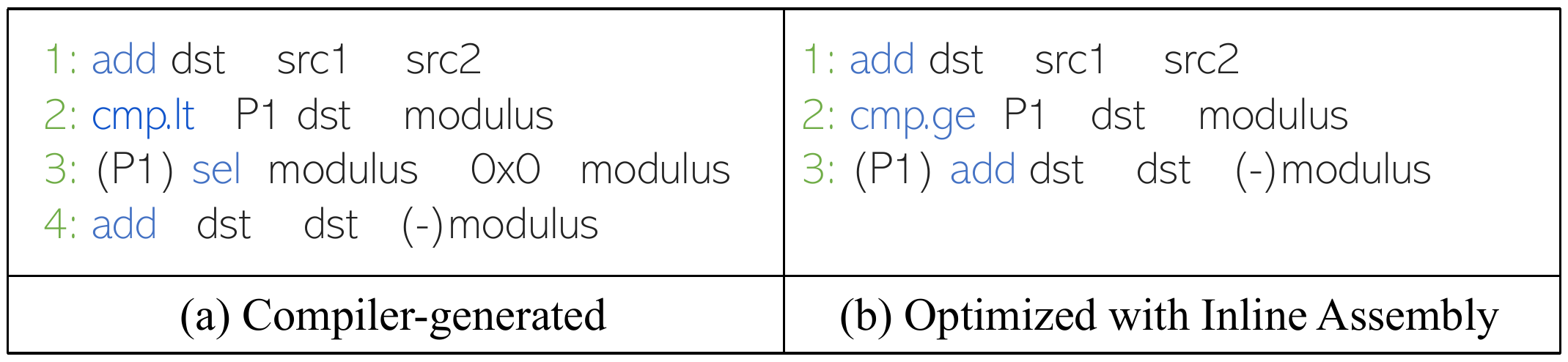}
\caption{Pseudo int64 addmod assembly}
\label{fig:inline-asm-add}
\end{figure}

The optimization for this operation is simple yet effective. The compiler-generated assembly sequence shown in Figure \ref{fig:inline-asm-add}(a) is replaced with the sequence shown in Figure \ref{fig:inline-asm-add}(b) eliminating one instruction from the sequence. It is important to note that even applying a small optimization to core operations at the assembly level enables direct benefits to other complex HE operations, since these core operations are ubiquitous in HE libraries.
\paragraph{Unsigned Integer Multiplication (mul64)}
Other cases at which inline assembly can be useful is when the compiler's generated code sequence does not exactly or efficiently match the required functionality and as a result, uses more expensive operations or code sequences.

%An example of this case was found in the compiler's generated code for our int64 multiplication implementation. A compilation deficiency related to variables' type-casting was discovered. By default, the compiler looks for ways to minimize the number of type-casting instructions. As this would seem like the right thing to do, it can also have a detrimental effect on the code, considering that operands' types play a significant role in how an instruction is implemented. For example, an integer multiplication where both operands are of type int32 can be executed with a single assembly instruction, compared to a lengthier, emulated multiplication implementation when the operands are of type int64.

An example of this case can be found in the compiler-generated code for int64 multiplication, which is a compilation deficiency related to variables' type-casting. By default, the compiler looks for ways to minimize the number of type-casting instructions. Although this would generally seem beneficial, it can be detrimental in some cases, considering that operands' types play a significant role in how an instruction is implemented. For example, an integer multiplication, where both operands are int32, can be executed with a single assembly instruction, compared to a lengthier, emulated multiplication implementation when the operands are of type int64.

This presents an example of how knowledge of the underlying instruction set architecture can provide very specific low-level optimization opportunities. We target one of the multiplication instructions available in our hardware for this purpose - we will refer to this instruction as \textit{mul\_low\_high}. The instruction receives two 32-bit operands and stores both low and high 32 bits of the result in a 64-bit destination\cite{mul64}.

% <<<< "to bypass this deficiency" <<<<
% yz: could you elaborate what deficiency you are referring to and why this is a deficiency? - I think "deficiency" here actually refers to "considering that operands' types play a significant role in how an instruction is implemented. For example, an integer multiplication, where both operands are int32, can be executed with a single assembly instruction, compared to a lengthier, emulated multiplication implementation when the operands are of type int64". However, this remains unclear for me to understand the exact compiler deficiency. --- this might be alleviated after you add the figure :)

\begin{figure}[ht]
\centering
\includegraphics[width=0.4\textwidth]{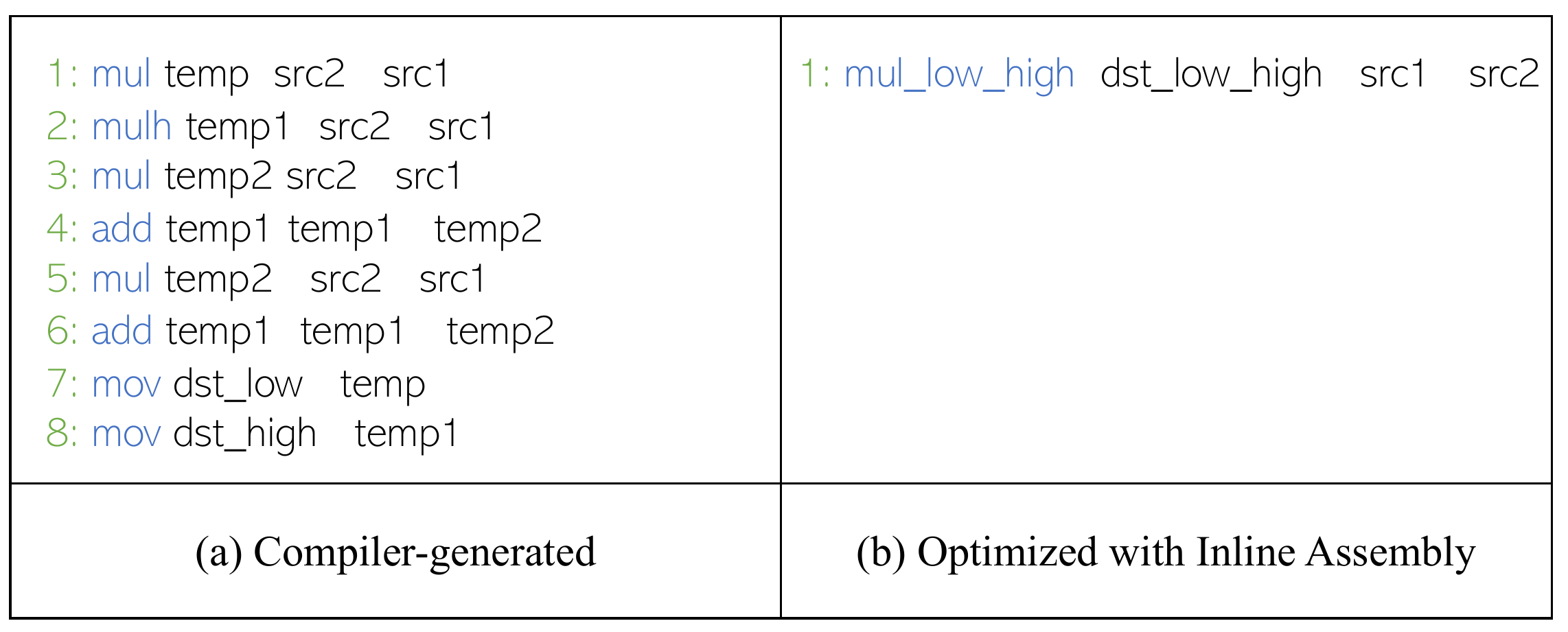}
\caption{Pseudo mul64 inline optimization}
\label{fig:inline-asm-mul}
\end{figure}

At every multiplication where the compiler eagerly optimizes a int32 operands to int64, we utilize inline assembly to bypass this deficiency by forcing the compiler to keep the appropriate types and also utilize the \textit{mul\_low\_high} instruction, as shown in Figure \ref{fig:inline-asm-mul} This optimization yields a ${\sim}60\%$ reduction in instruction count from our original int64 multiplication implementation.

% As mentioned previously, applying low-level optimizations to core operations directly impact our algorithm and kernel performance. Table \ref{kernel_count_ops} shows an example of some NTT/iNTT kernels and the number of \textit{mul64} operations in each kernel. Hence, optimizations aimed at our arithmetic operations will greatly impact the performance as will be discussed in more detail in the Evaluation Section.
As mentioned previously, applying low-level optimizations to core operations directly impact our algorithm and kernel performance. Optimizations aimed at our core arithmetic operations will greatly impact the performance of HE as will be discussed in more detail in Section \ref{sec:results}.

% \begin{table}[ht]
% \centering
% \caption{mul64 count in example NTT/iNTT kernels}
% \label{kernel_count_ops}
% \begin{tabular}{|l|c|}
% \hline
% \textbf{Kernel}               & \multicolumn{1}{l|}{\textbf{mul64 ops}} \\ \hline
% RnsInvDwtGapLocalRescaleFused & 47                                      \\ \hline
% RnsInvDwtGapLocalFused        & 47                                      \\ \hline
% RnsInvDwtGapLocalRelinFused   & 47                                      \\ \hline
% RnsDwtGapLocalRadix8          & 24                                      \\ \hline
% RnsInvDwtGapLocalRelin        & 16                                      \\ \hline
% RnsDwtGapLocal                & 16                                      \\ \hline
% \end{tabular}
% \end{table}

\subsection{Algorithmic level optimizations (NTT)}

\begin{figure}[ht] \centering
\subfigure[{Profiling on Device1}]
{
\includegraphics[width=0.215\textwidth]{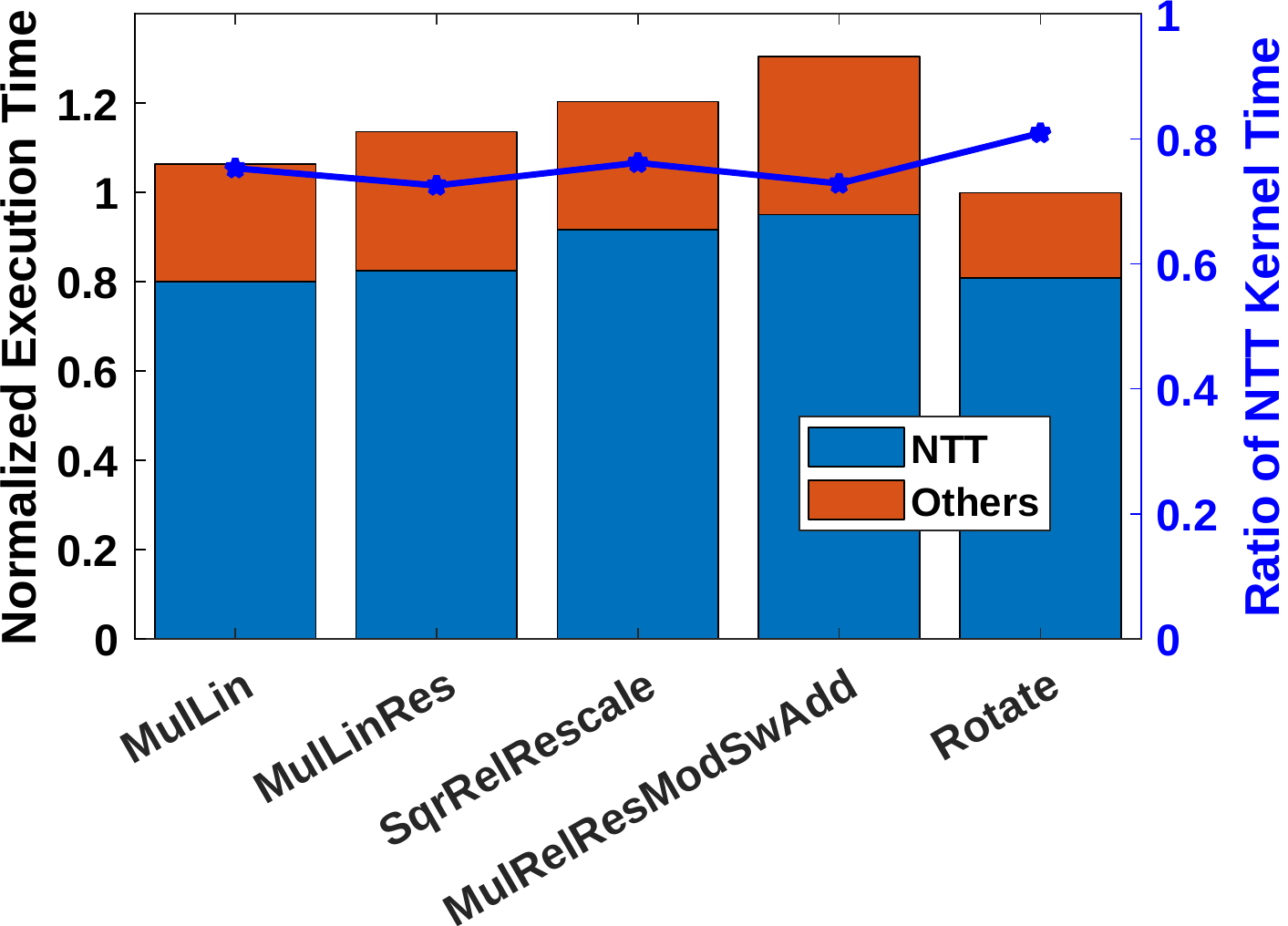}
}
\hspace{-3mm}
\vspace{-3mm}
\subfigure[Profiling on Device2]
{
\includegraphics[width=0.215\textwidth]{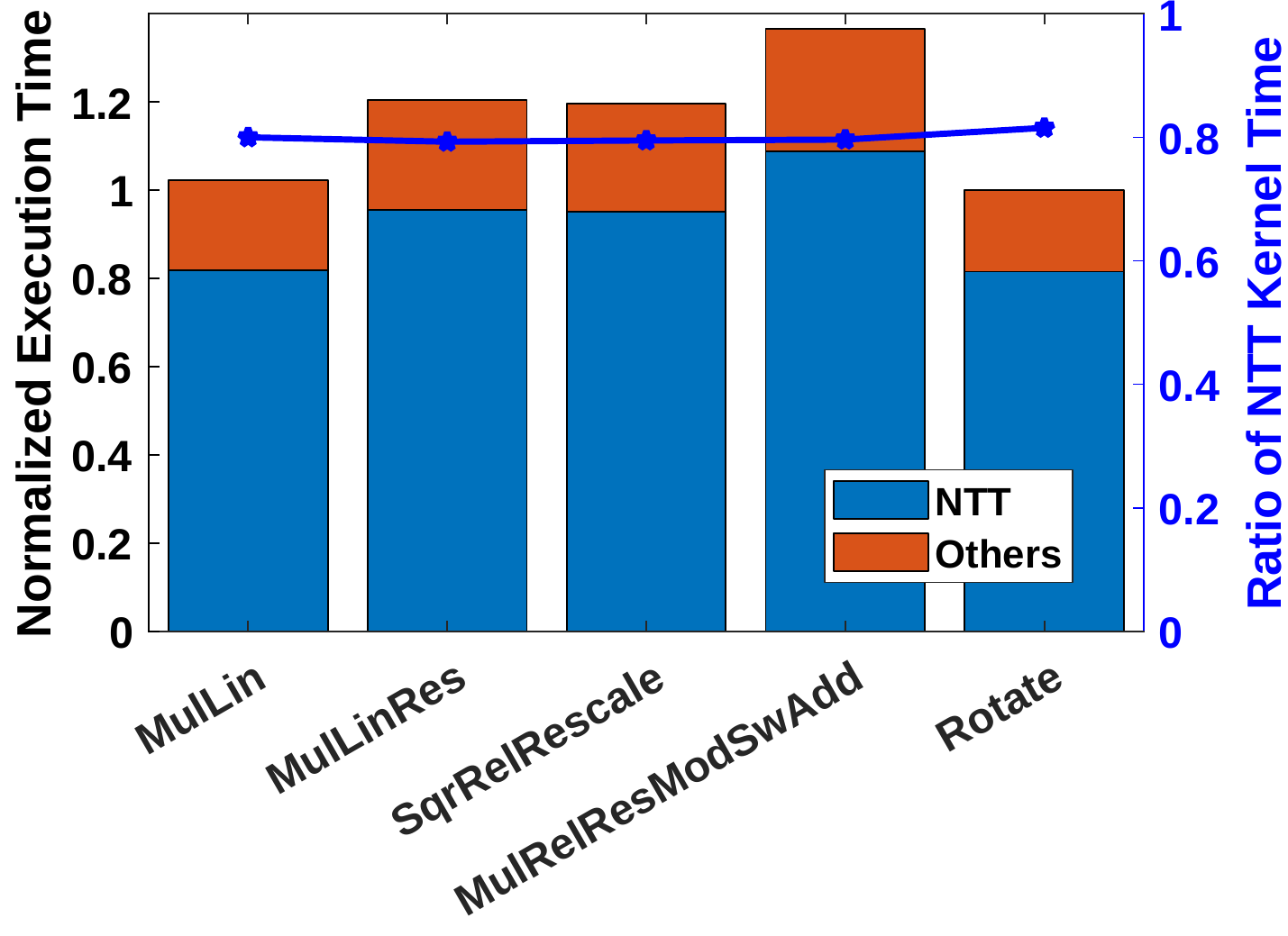}
}
\caption{Profiling for HE routines}
\label{fig:ntt-profiling}
\end{figure}

An efficient NTT implementation is crucial for HE computations since it accounts for a substantial percentage of the total HE computation time\cite{roy2019fpga,jung2020heaan,kim2020accelerating}. Figure \ref{fig:ntt-profiling} presents the relative execution time of five HE evaluation routines and the percentage of NTT in each routine that we benchmarked on two latest Intel GPUs, Device1 and Device2. We observe that NTT accounts for 79.99\% and 75.64\% of the total execution time in average on these two platforms. The profiling confirms the significance of optimizing NTT for HE on Intel GPUs.

%Different from prior NTT works \cite{kim2020accelerating, cuFHE}, where the hierarchical NTT implementation is adopted, we present a staged implementation of NTT to leverage the hierarchical memory architecture on Intel GPUs --- global memory, shared local memory and general-purpose register file.

\subsubsection{Naive radix-2 NTT} 
\begin{figure}[ht]
\centering
\includegraphics[width=0.46\textwidth]{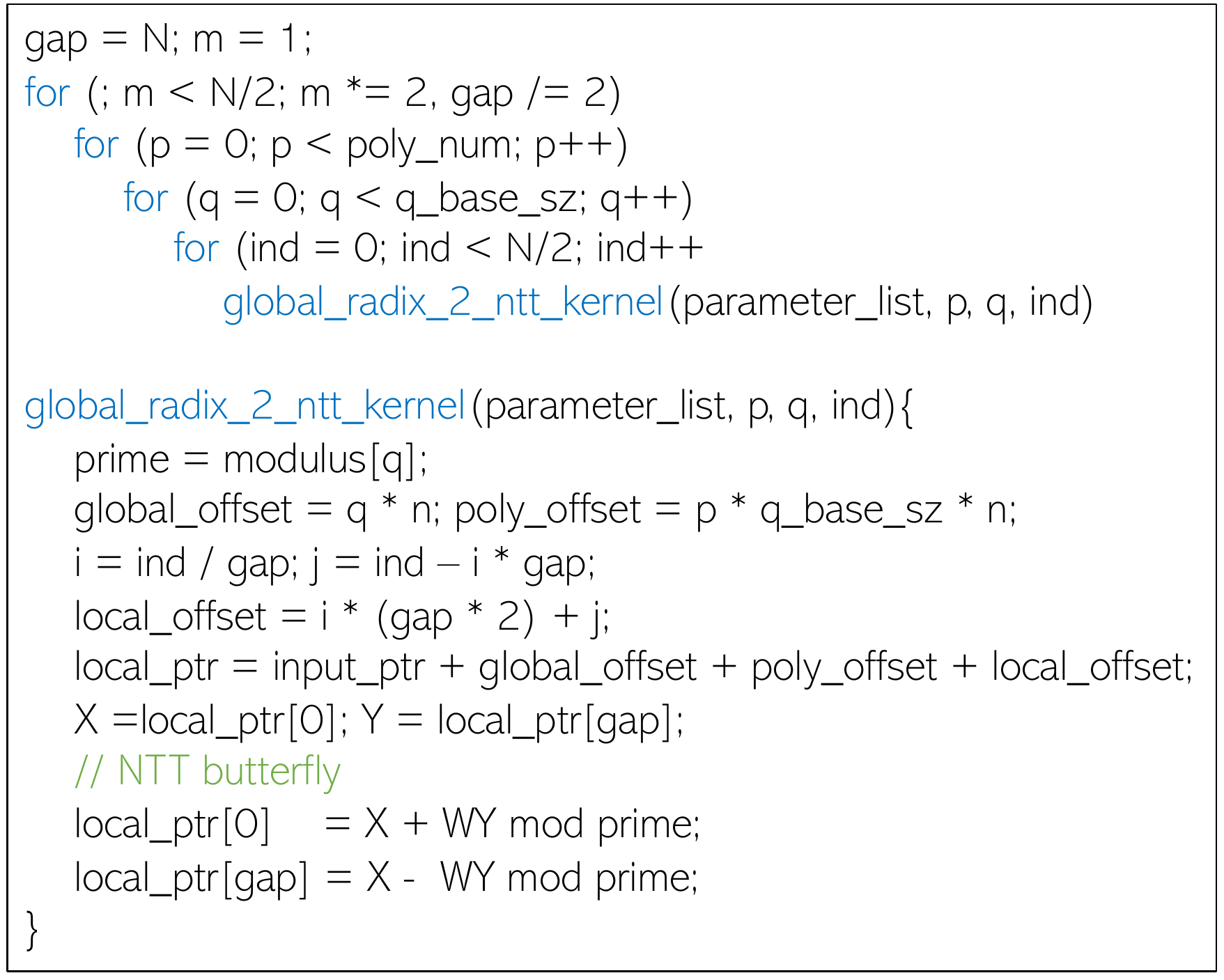}
\caption{Naive NTT pseudo code}
\label{fig:ntt-naive}
\end{figure}

We start NTT optimizations from the most naive radix-2 implementation. This reference implementation of NTT, as shown in Figure \ref{fig:ntt-naive}, distributes rounds of radix-2 NTT butterfly operations among work items. In each round of NTT computation, all the work items compute their own butterfly operation and exchange data with other work-items using global memory. More specifically, the $k$-th element will exchange with the $k+gap$ th element, while the exchanging gap sizes are halved after each round of NTT until it becomes equal to $1$. Accordingly, the $m$-loop in Figure \ref{fig:ntt-naive} is executed $\log(N)$ times throughout the NTT computation, where for each iteration of the $m$-loop, one accesses the global memory $2N$ times. Here we multiply it by two because of both load and store operations. We ignore the twiddle factor memory access number in this semi-quantitative analysis.

At the lowest level, the NTT butterfly computation is accelerated using Algorithm \ref{fig:ntt-butterfly}\cite{harvey2014faster}. Since the output $X',Y'$ of Algorithm \ref{fig:ntt-butterfly} are both in $[0,4p)$, to ensure all elements of the output NTT sequence falls inside of the interval $[0,p)$, a last round offsetting needs to be appended to the end of NTT computations. Therefore, the naive implementation of an $N$-point NTT needs to access the global memory $2N\log(N)$ times for the NTT and $2N$ extra times for last round processing. A three-dimensional DPC++ kernel corresponding to the three layers of the for loop is invoked when deploying this algorithm on Intel GPUs. This kernel reaches only 10\% of the peak performance for a 32K-point, 8-RNS-size, 1024-instance NTT.

\begin{algorithm}[ht]
  \caption{Optimized NTT butterfly computation}
  \textbf{Input:} $0\leq X,Y\leq 4p$, $p < \beta/4$, $0<W<p, 0<\left \lfloor W\beta /p \right \rfloor=W'<\beta$\;
  \textbf{Output:}\newline 
  $X'=X+WY \mod p$\; 
  \ \ $Y'=X-WY \mod p$\;
  \ \ $0\leq X', Y'\leq 4p$\;
  \algorithmicif $\ X\geqslant 2p$ \algorithmicthen{\ $X\leftarrow X - 2p$}\;
  $Q\leftarrow \left \lfloor W'Y/\beta \right \rfloor$\;
  $T\leftarrow ( WY - Qp) \mod \beta$\;
  $X'\leftarrow X + T $\;
  $Y'\leftarrow X - T + 2p $\;
  \algorithmicreturn{\ X', Y'}
  \label{fig:ntt-butterfly}
\end{algorithm}

\subsubsection{Staged radix-2 NTT with shared local memory}

Due to the low bandwidth and high latency of the global memory, the operational density of naive radix-2 NTT is far from optimal. Therefore, its performance is bounded by the global memory bandwidth. To address this issue, we keep data close to computing units by leveraging shared local memory (SLM) in Intel GPUs, a memory region that is accessible to all the work-items belonging to the same work-group. 

Given that the data exchanging gap size are halved after each round of NTT, there exists a certain round when the gap size becomes sufficiently small so that all data to exchange can be held in SLM. We call this threshold gap size \verb|TER_SLM_GAP_SZ|, after which we retain the data in SLM for communication between work-items to avoid the expensive global memory latency. For example, in a 32K-point NTT, we first compute three rounds of NTT and exchange data using global memory and then the data exchanging gap size has decreased to 4K. We set the \verb|TER_SLM_GAP_SZ| to 4K because the size of the SLM on most Intel GPUs is 64KB, which can hold 8K int64 elements. For the remaining rounds, the data are held in SLM until all computations are completed.

\subsubsection{SIMD shuffling}

\begin{figure}[ht]
\centering
\includegraphics[width=0.36\textwidth]{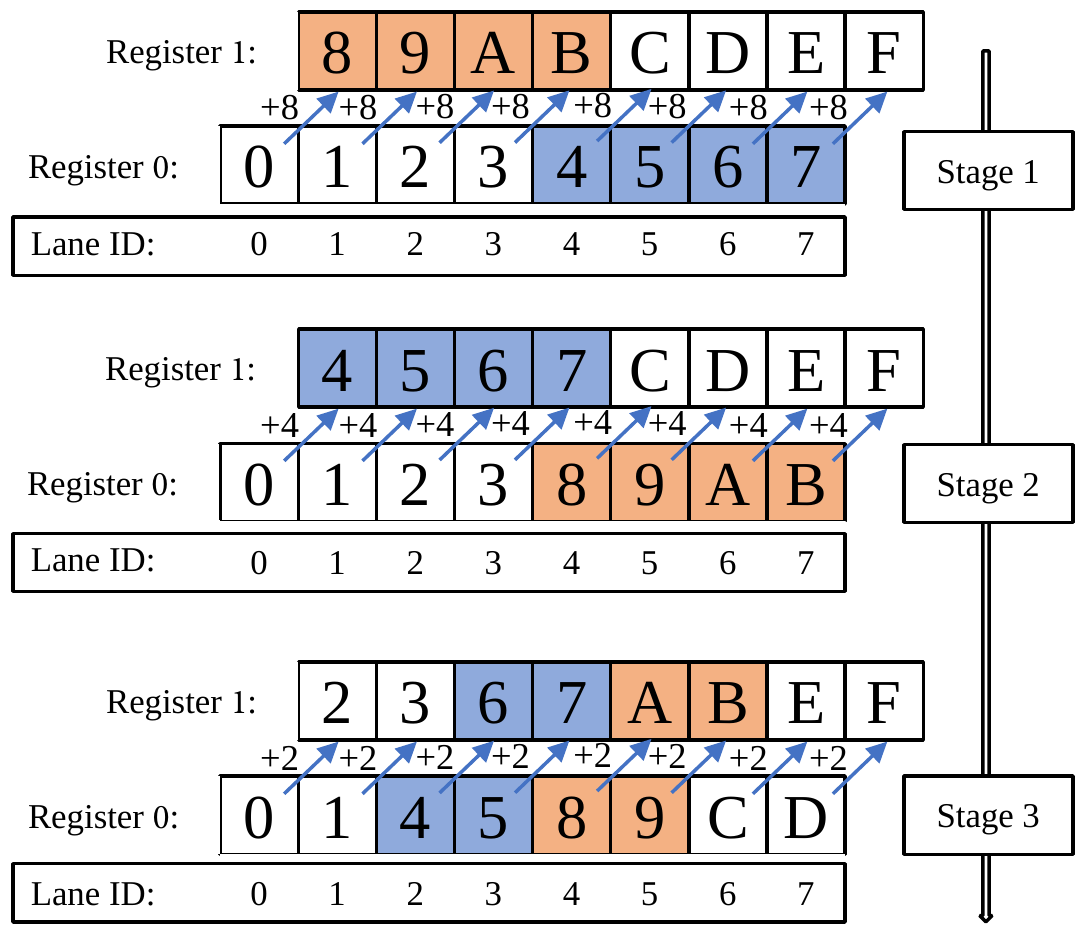}
\caption{SIMD shuffling for data exchanging in NTT.}
\label{fig:ntt-simd-shuffling}
\end{figure}

In addition to introducing shared local memory, when the exchanging distance becomes sufficiently small that all data to exchange are held by work-items in the same subgroup, we perform SIMD shuffling directly among all the work-items in the same subgroup after NTT butterfly computations. In Figure \ref{fig:ntt-simd-shuffling}, we present the rationale of two SIMD shuffling operations among three stages. When the SIMD width equals to 8, there are 8 work-items in a subgroup. For the radix-2 NTT implementation, each work-item holds two elements of the NTT sequence in registers, namely one slot. We denote two local registers of each work-item as Register 0 and Register 1. At the end of Stage 1, one needs to exchange data at positions ``8, 9, A, B" with ``4, 5, 6, 7". Such operations can be implemented using \verb|shuffle| of the Intel extension of DPC++ \cite{llvm-intel-shuffle}. That being said, four lanes (lane ID: 0, 1, 2, 3) are exchanging data stored in their Register 1 with Register 0 of lane 4, 5, 6, 7. At the end of Stage 2, where the exchanging gap size is halved from 8 to 4, lanes 0, 1 will exchange data of their Register 1 with Register 0 of lanes 2, 3; similarly, lanes 4, 5 exchange their Register 0 with Register 1 of lanes 6, 7.

\begin{figure}[ht]
\centering
\includegraphics[width=0.46\textwidth]{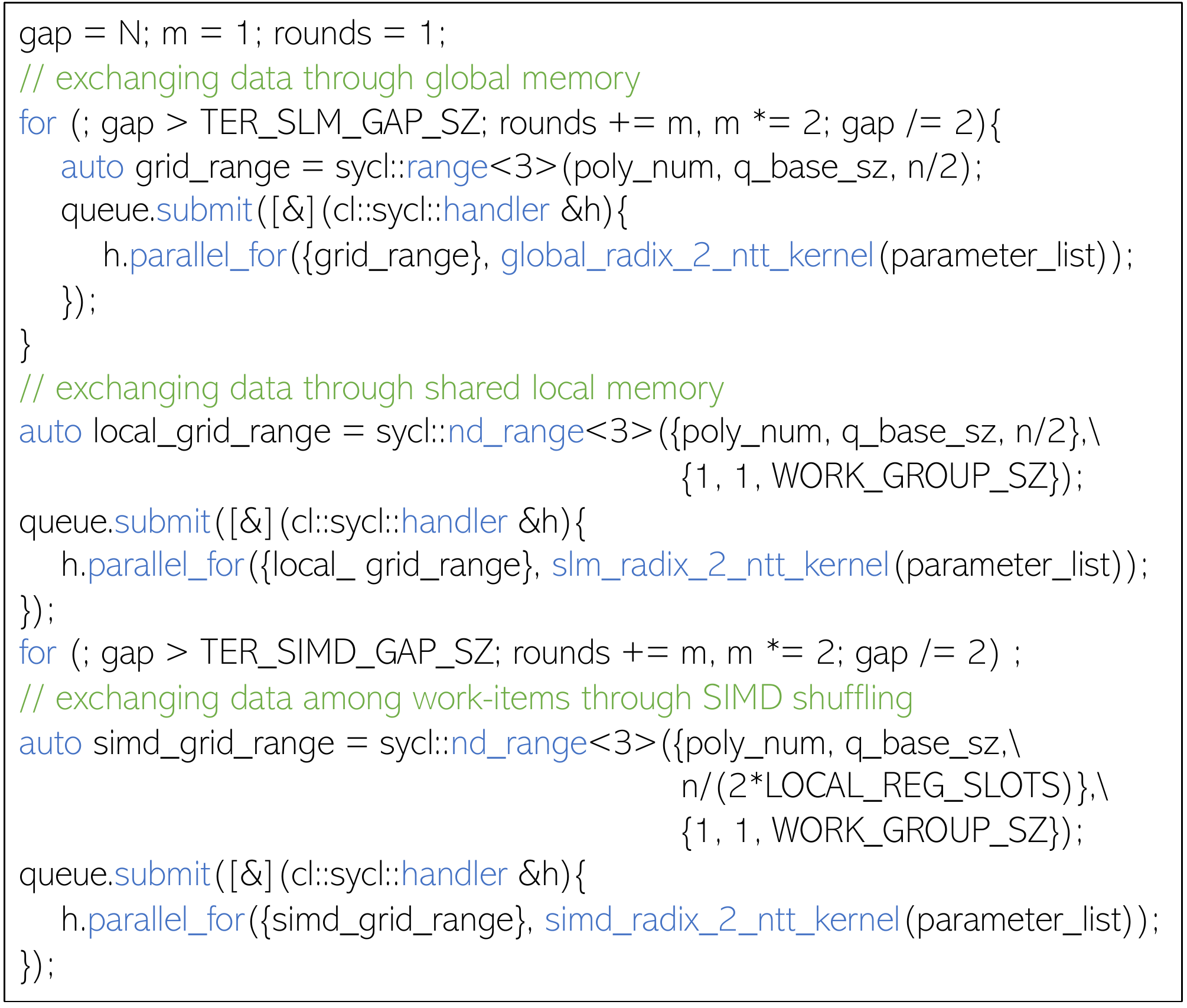}
\caption{Staged NTT pseudo code}
\label{fig:ntt-staged}
\end{figure}

Figure \ref{fig:ntt-staged} summarizes our staged NTT implementation with both SLM and SIMD shuffling considered. Before the gap size reaches the threshold to exchange data in SLM, \verb|TER_SLM_GAP_SZ| work-items communicate through the global memory by invoking \textit{global\_radix\_2\_ntt\_kernel}. After reaching the SLM threshold, one computes NTT butterfly operations and exchanges data through SLM by calling \textit{slm\_radix\_2\_ntt\_kernel} until the gap size equals the threshold to exchange data using SIMD shuffling inside subgroups, which is handled by the kernel \textit{simd\_radix\_2\_ntt\_kernel}. It is worth mentioning that the SIMD kernel is fused with the aforementioned last round processing operation to reduce all NTT elements to $[0,p)$.

\begin{figure}[ht]
\centering
\includegraphics[width=0.46\textwidth]{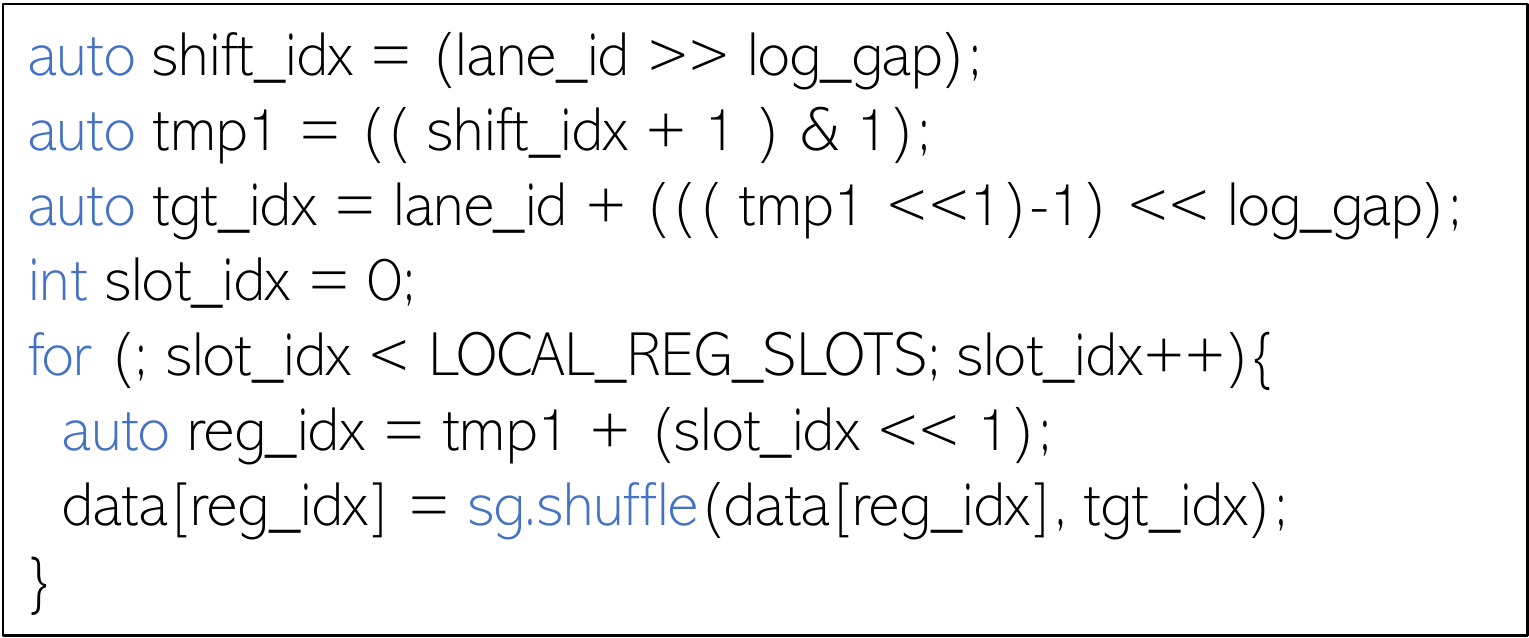}
\caption{NTT SIMD shuffling pseudo code}
\label{fig:ntt-simd-shuffling-pseudo}
\end{figure}

\subsubsection{More aggressive register blocking}

Intel GPUs typically consist of 4KB GRF for each EU thread. When the SIMD width equals 8, that indicates 8 work-items are bounded executing as an EU thread in the SIMD manner. For the radix-2 NTT implementation, each work-item needs four registers, where two of them are used to hold NTT data elements and the other two are for twiddle factors. Therefore, the NTT-related computation consumes $4\cdot 8\cdot 8$B$ = 256$B GRF for each EU thread --- 6.25\% of total GRF, indicating that the hardware is significantly underutilized at the register level. Rather than initializing only one slot of registers, one can assign more workloads to each work-item. For example, we can allocate 4 slots of registers for each work-item to hold NTT element data for each work-item. For a subgroup of size 8, there are $8\cdot 4=32$ NTT elements being held in registers in the SIMD kernel. This is referred as SIMD(32,8) in following discussions.

Figure \ref{fig:ntt-simd-shuffling-pseudo} shows the pseudo code of the multi-slot SIMD shuffling operation. When \verb|LOCAL_REG_SLOTS| is set to 1, this pseudo code degrades to the single-slot implementation. Compared with the single-slot implementation, the multi-slot SIMD implementation results in fewer accesses to shared local memory, but suffers from higher register pressure and the in-register data exchange overhead. In practice, the efficiencies of both 2-slot SIMD(16,8) and 4-slot SIMD(32,8) implementations are worse than the 1-slot SIMD(8,8), suggesting that negative aspects dominate the performance.

\subsubsection{High-radix NTT}

\begin{comment}

\begin{figure}[ht]
\centering
\includegraphics[width=0.4\textwidth]{figures/NTT_new+6_cropped.pdf}
\caption{Radix-4 16-point NTT}
\label{fig:radix-4}
\end{figure}

\end{comment}

Prior optimization trials indicate that the cost of data exchanging can offset the acceleration from introducing SLM. With this observation in mind, we seek optimizations to directly reduce the number of data exchanges and communication among work-items so that we can reduce the number of SIMD shuffling and memory accesses. Therefore, we adopt high-radix NTT implementations. Compared with the radix-2 implementation, a high-radix NTT aggregates more data elements into a group, and performs calculations with all the data held in registers. 

We use radix-8 as an example to demonstrate high-radix NTT algorithms. Each work-item allocates 8 registers to hold NTT elements and 8 more registers to hold root power and root power quotients as for the twiddle factors. For a specific round where the exchanging gap size is $gap$, one loads eight NTT elements from either global or shared local memory, indexing at $\{k,k+gap,k+2\cdot gap,...,k+7\cdot gap\}$. There are three internal rounds of NTT for each radix-8 kernel. In the first internal round, four pairs of 2-point butterfly computations are performed among $\{x[k],x[k+4\cdot gap]\}$, $\{x[k+gap],x[k+5\cdot gap]\}$, $\{x[k+2\cdot gap],x[k+6\cdot gap]\}$ and $\{x[k+3\cdot gap],x[k+7\cdot gap]\}$. For the second round, these eight elements, still held in registers, are re-paired to $\{x[k],x[k+2\cdot gap]\}$, $\{x[k+gap],x[k+3\cdot gap]\}$, $\{x[k+4\cdot gap],x[k+6\cdot gap]\}$ and $\{x[k+5\cdot gap],x[k+7\cdot gap]\}$ so that Algorithm \ref{fig:ntt-butterfly} can be leveraged. In the last internal round of the radix-8 kernel, each two consecutive elements are paired and fed into the 2-point butterfly algorithm. After all in-register computations are completed, we store results back to either global memory or shared local memory, depending on whether it is a global memory kernel or a shared local memory kernel. Last round processing is fused with shared local memory high-radix NTT kernels.

% Therefore, registers are more aggressively re-used for high-radix NTT and less global memory access is required accordingly because now we only need $\log_4(N)$ rounds processing rather that the previous $\log_2(N)$ rounds. Although introducing such advantages, one cannot infinitely increase the radix because the GRF size in each EU is limited. Because the default SIMD width is 8, indicating that 8 work-items are bounded into an EU thread during execution. As discussed before, each EU thread consists of 4KB GRF. Since both NTT sequence elements and root power or root power quotients are \verb|uint64|, which counts for 8 Bytes, one can assign up to radix-8 implementation for each EU thread.

\begin{figure}[ht]
\centering
\includegraphics[width=0.3\textwidth]{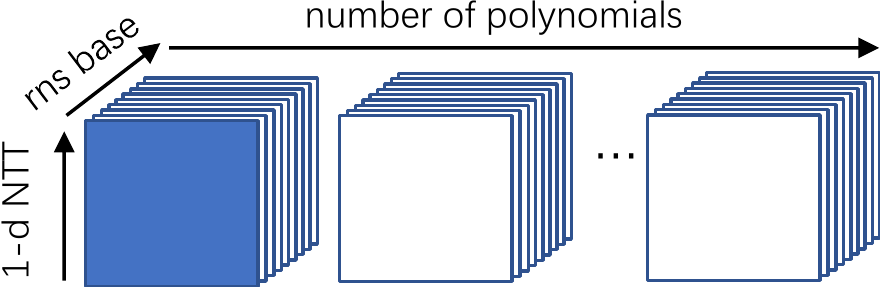}
\caption{The parallelism of NTT for HE}
\label{fig:ntt-parallelism}
\end{figure}

Figure \ref{fig:ntt-parallelism} shows the parallelism of the staged NTT implementation. For a 32K-point NTT shared local memory kernel, we assign 4K NTT elements to each work-group. Therefore, 8 work-groups, which are mapped to SubSlices, are activated. Meanwhile, both RNS and the batched number of polynomials can provide us with additional parallelisms since the RNS size can be up to several dozens \cite{kim2020accelerating} while the batch size can be up to tens of thousands in real-world deep learning tasks\cite{keskar2016large}.

\subsection{Application-level Optimizations}

In addition to instruction-level and algorithm-level optimizations, we also optimize our GPU-accelerated HE library from the application level.

\subsubsection{Memory cache}
\begin{figure}[ht]
\centering
\includegraphics[width=0.46\textwidth]{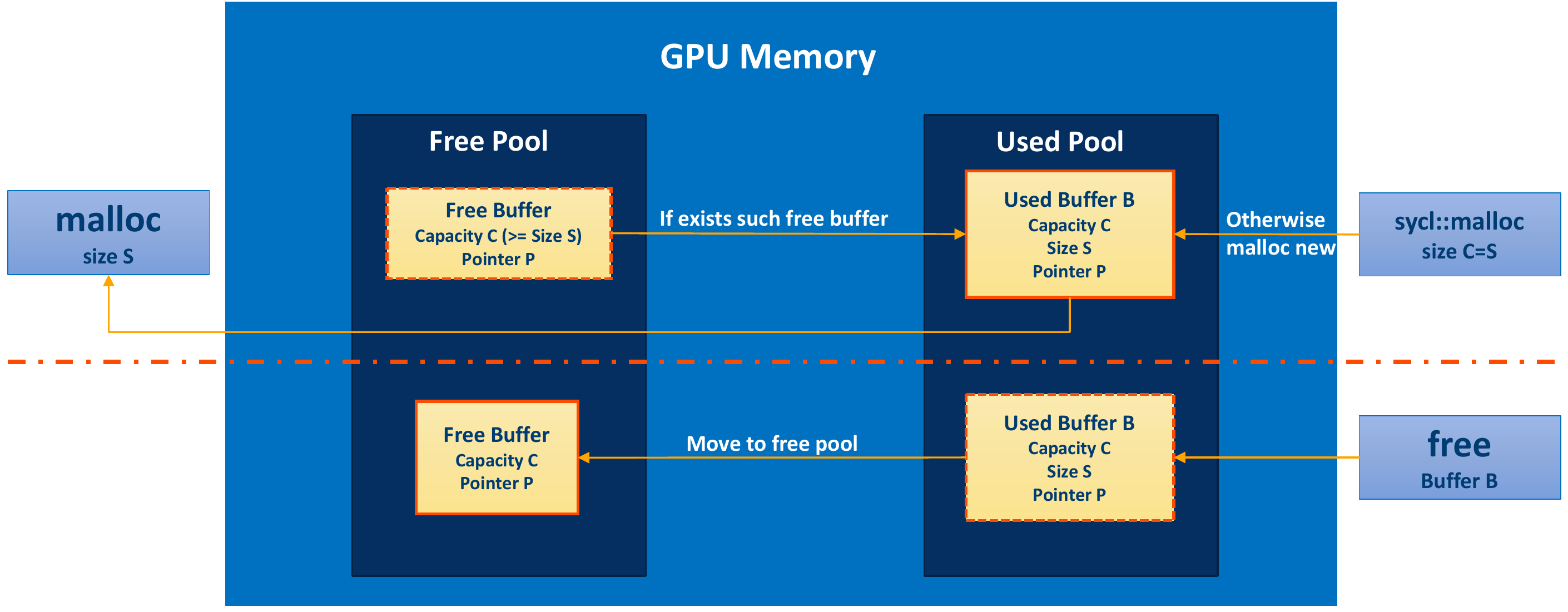}
\caption{Memory cache design}
\label{fig:mem-cache}
\end{figure}

To reduce the overhead introduced by runtime memory allocation, we design a memory cache mechanism to our HE library, as shown in Figure \ref{fig:mem-cache}. Similar to Microsoft SEAL, we introduce a memory pool to reuse allocated GPU memory buffers in the HE pipeline. A request for a new GPU memory buffer is routed through the memory cache for any existing free buffer with a capacity larger than the current request. If such buffer is found, this existing buffer is reused instead of allocating a new one. Upon freeing such a buffer, it is moved back to the free pool for potential reuse.

\subsubsection{Inter-device scaling} At this time, DPC++ does not implicitly support the multi-tile submission. As such, the workloads cannot automatically be distributed over all the computing units of a multi-tile Intel GPU. Therefore, we explicitly submit workloads to the multi-tile device through multiple queues. We refer a reader to \cite{dpcmultiqueue} for more details of the DPC++ multi-queue implementation.
%\subsection{Asynchronous execution} 

\begin{comment}

\begin{figure*}
\centering
\includegraphics[width=1.0\textwidth]{figures/NTT_new+4_cropped.pdf}
\caption{Software pipelining design. \small{{\it \textmd{Each letter represents a vectorized instruction. L: Load; M1: Mul; M2: Duplicated Mul; C: Vectorized Comparison; S: Store; BS: Checkpoint original value before scaling into an unused register, then Store the computing result back to memory; R: Restore from a checkpoint register.}}}}
\end{figure*}

\end{comment}

\section{Evaluation} \label{sec:results}

%We evaluate our optimizations on two engineering sample Intel GPUs with the latest microarchitecture. Due to confidentiality requirements, at this time, we do not disclose hardware specifications of these GPUs. For the same purpose, we present performance data by showing normalized execution time rather than the absolute elapsed time or showing performance in the unit of GFLOPS. The first Intel GPU, denoted as Device1 in following discussions, is a multi-tile GPU while the second Intel GPU, Device2, is a single-tile GPU. We utilize up to 2 tiles in the multi-tile Device1 for performance benchmarking and efficiency estimation. Both GPU devices are connected with 24-core Intel Icelake server CPUs, whose boost frequency is up to 4 GHz. The associated main memory systems are both 128GB at 3200 MHz. We compile programs using Intel Data Parallel C++ (DPC++) Compiler \verb|2021.3.0| with the optimization flag \verb|-O3|.

We evaluate our optimizations on two Intel GPUs with the latest microarchitecture. Due to confidentiality requirements, at this time, we do not disclose hardware specifications of these GPUs. For the same purpose, we present performance data by showing normalized execution time rather than the absolute elapsed time or showing performance in the unit of GFLOPS. The first Intel GPU, denoted as Device1 in following discussions, is a multi-tile GPU while the second Intel GPU, Device2, is a single-tile GPU. We utilize up to 2 tiles in the multi-tile Device1 for performance benchmarking and efficiency estimation. Both GPU devices are connected with 24-core Intel Icelake server CPUs, whose boost frequency is up to 4 GHz. The associated main memory systems are both 128GB at 3200 MHz. We compile programs using Intel Data Parallel C++ (DPC++) Compiler \verb|2021.3.0| with the optimization flag \verb|-O3|.

\subsection{Optimizing NTT on Intel GPUs}

Figure \ref{fig:ntt-profiling} shows that NTT accounts for at least 70\% of the total execution time of HE evaluations routines on Intel GPUs. Therefore, we start optimizations from this decisive algorithm.

\subsubsection{First Trial: optimizing NTT using SLM and SIMD}

\begin{figure}[ht] \centering
\vspace{-2mm}
% \hspace{-1mm}
\subfigure[{Speedup}]
{
\includegraphics[width=0.215\textwidth]{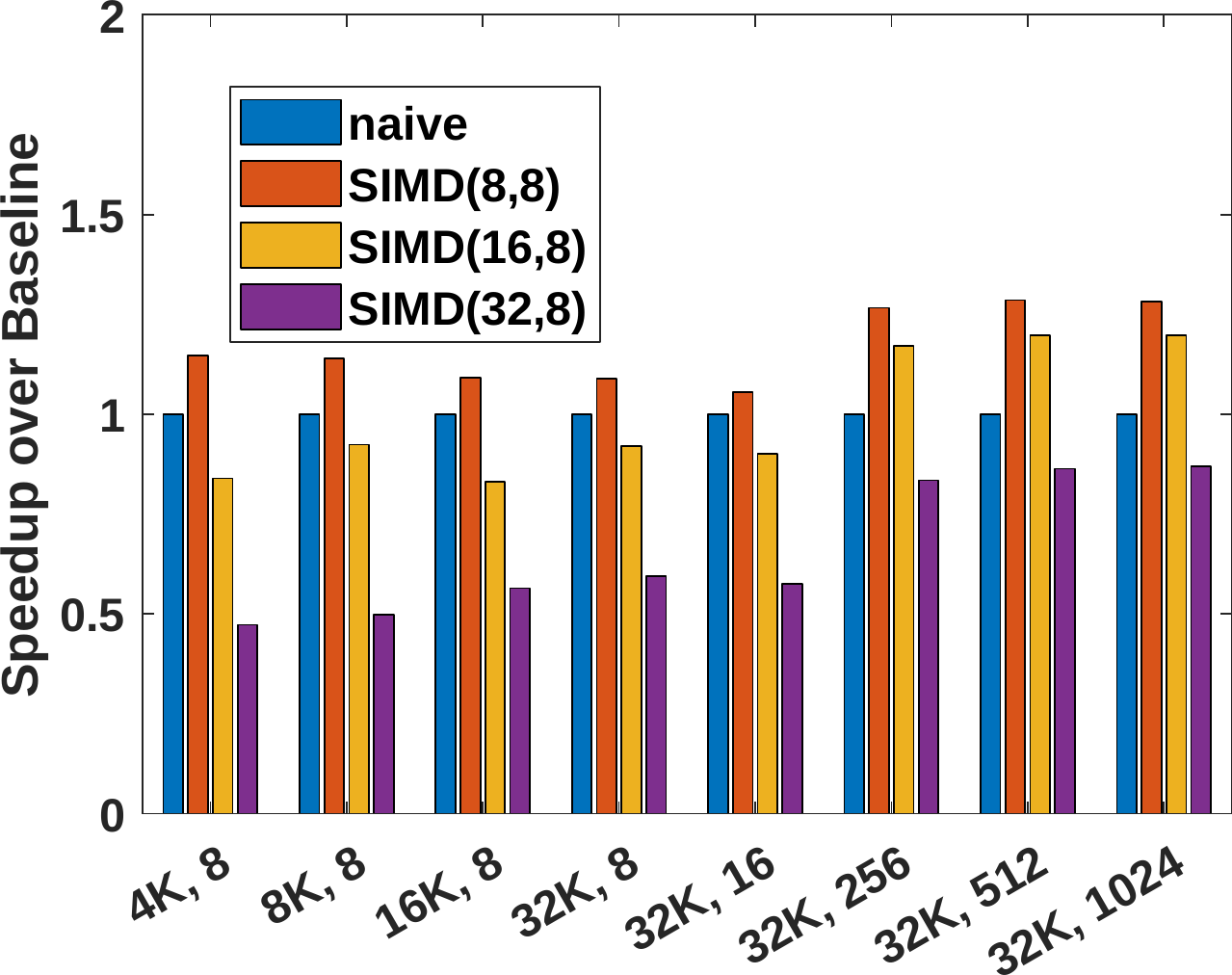}
}
\hspace{-3mm}
\vspace{-1mm}
\subfigure[Efficiency of 32K-point NTT]
{
\includegraphics[width=0.215\textwidth]{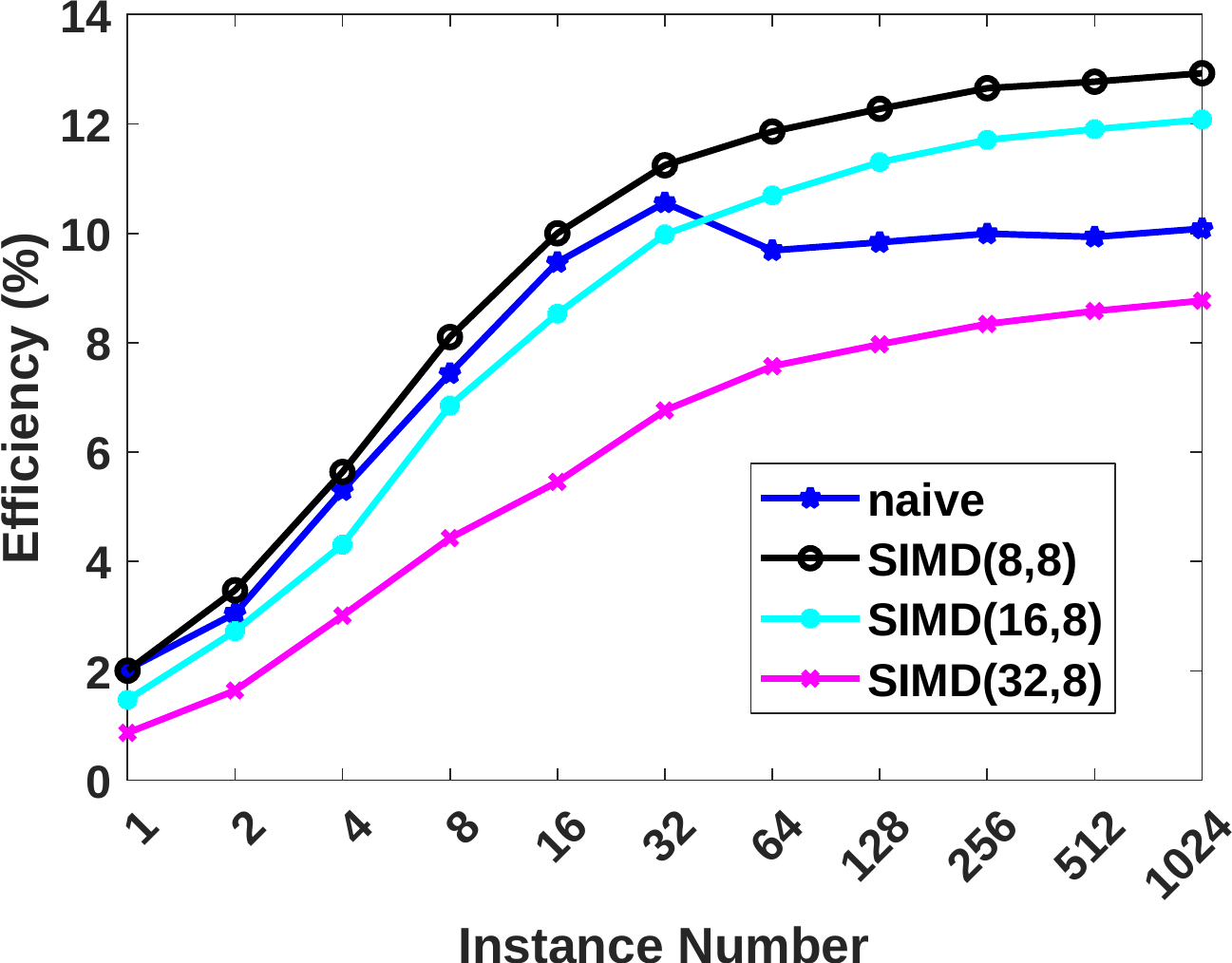}
}
\vspace{-1mm}
\caption{Radix-2 NTT with SLM and SIMD on Device1}
\label{fig:ntt-simd-bench}
\end{figure}

Figure \ref{fig:ntt-simd-bench} (a) compares the speedup of our first batch of NTT trials using the staged NTT implementation over the naive GPU implementation of NTT described in Figure \ref{fig:ntt-naive}. We use SIMD(\verb|TER_SIMD_GAP_SZ|,\verb|SIMD_WIDTH|) to denote the different implementation variants. The number of register slots for each work-item can be computed by dividing \verb|TER_SIMD_GAP_SZ| over \verb|SIMD_WIDTH|. For example, each work-item holds a pair of NTT elements in registers for SIMD(8,8), and 4 pairs of NTT elements for SIMD(32,8). With the shared local memory as well as the SIMD shuffling for data exchanging among work-items included, we observe that SIMD(8,8) is faster than the baseline by up to 28\%. Meanwhile, SIMD(16,8) is slightly slowed down compared with SIMD(8,8) but remains up to 19\% faster than baseline. This indicates that the non-negligible cost of SIMD shuffling leads to the unfavorable performance. Accordingly, SIMD(32,8), which more aggressively performs SIMD shuffling and in-register data exchange than previous two variants, becomes even slower than the baseline.

Figure \ref{fig:ntt-simd-bench} (a) compares the efficiency of each NTT variant on Device1. The efficiency is computed by dividing the performance of each NTT implementation over the computed int64 peak performance, both in the unit of GFLOPS. The naive NTT reaches only 10.08\% of the peak performance for a 32K-point NTT with 1024 instances executed simultaneously. The best one, SIMD(8,8) obtains an efficiency up to 12.93\%. Since SIMD shuffling together with the SLM data communication fail to provide us with a high efficiency, we deduce that the cost of data communication is so high that radix-2 NTT cannot fully utilize the device.

\subsubsection{Second Trial: optimizing high-radix NTT using SLM}

\begin{figure}[ht] \centering
\vspace{-2mm}
% \hspace{-1mm}
\subfigure[{Speedup}]
{
\includegraphics[width=0.215\textwidth]{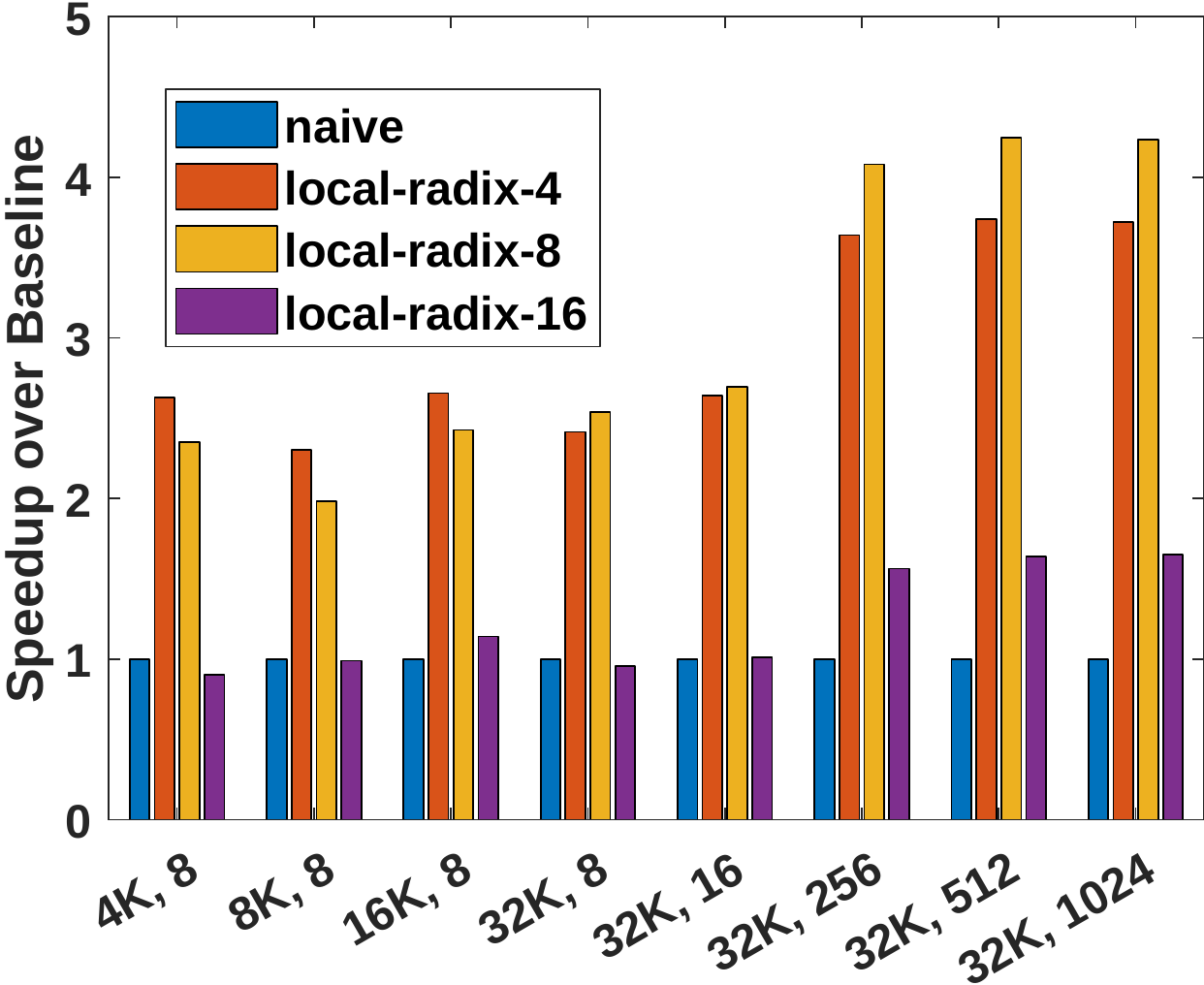}
}
\hspace{-3mm}
\vspace{-1mm}
\subfigure[Efficiency]
{
\includegraphics[width=0.215\textwidth]{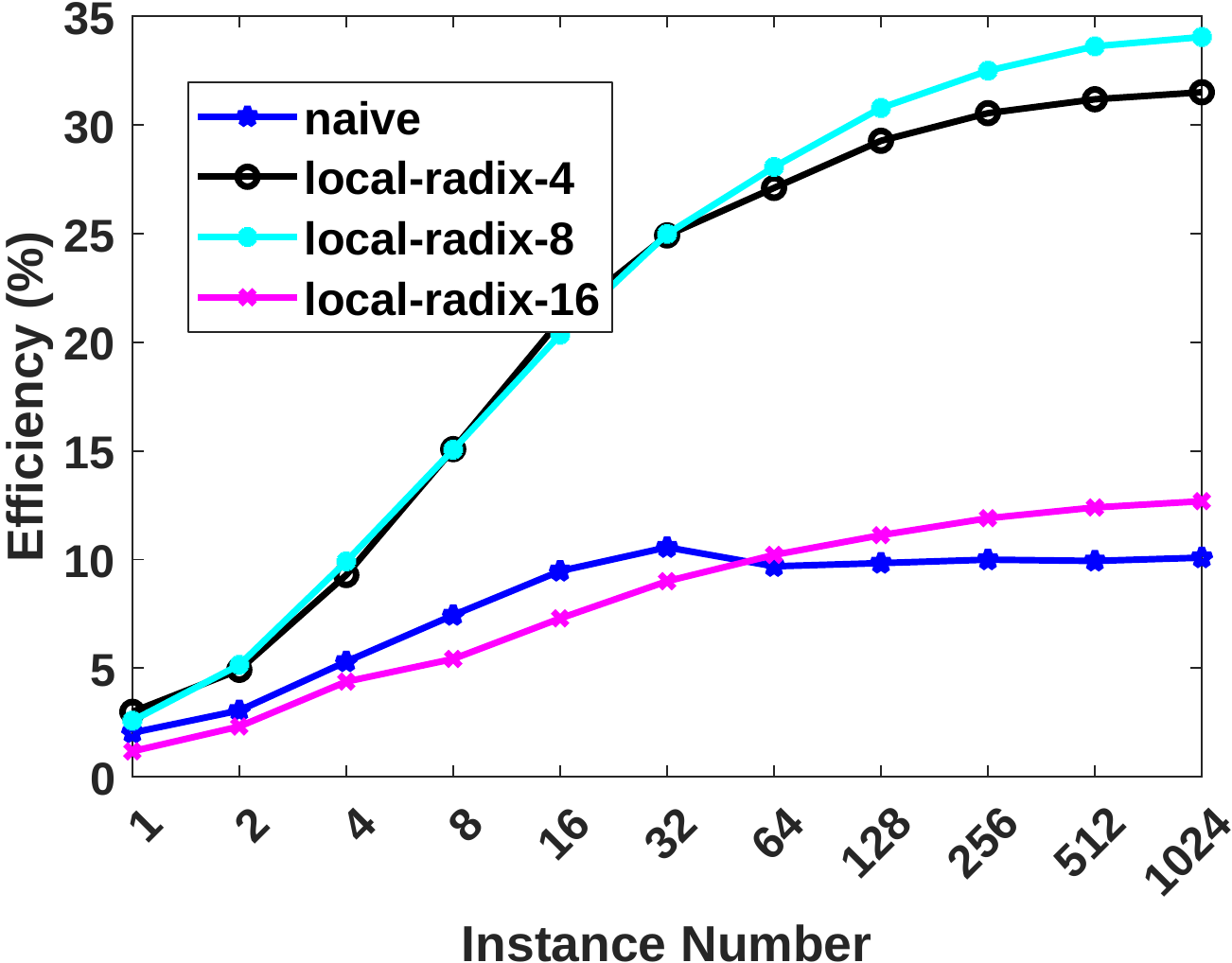}
}
%\vspace{-1mm}
\caption{High-radix NTT with SLM on Device1}
\label{fig:ntt-radix-bench}
\end{figure}

Figure \ref{fig:ntt-radix-bench} (a) compares the speedup of high-radix NTT implementations with shared local memory against the naive GPU baseline. High-radix NTT implementations re-use more data at the register-level, reducing the communication among work-items through either global memory or SLM. With the shared local memory also included, this time we obtain an up to 4.23X acceleration over the naive baseline. In Figure \ref{fig:ntt-radix-bench} (b), we see that the efficiency reaches its optimum, 34.1\% of the peak performance, at radix-8 NTT with 1024 instances instantiated for 32K-point NTT computations. The radix-16 NTT, though it brings more aggressive register-level data re-use and requires less data exchange among work-items, leads to the register spilling issue so its performance becomes significantly slower than radix-8 NTT.

\subsubsection{Assembly-level optimizations - add\_mod/mul\_mod}

\begin{figure}[ht] \centering
\vspace{-2mm}
% \hspace{-1mm}
\subfigure[{NTT with inline assembly}]
{
\includegraphics[width=0.215\textwidth]{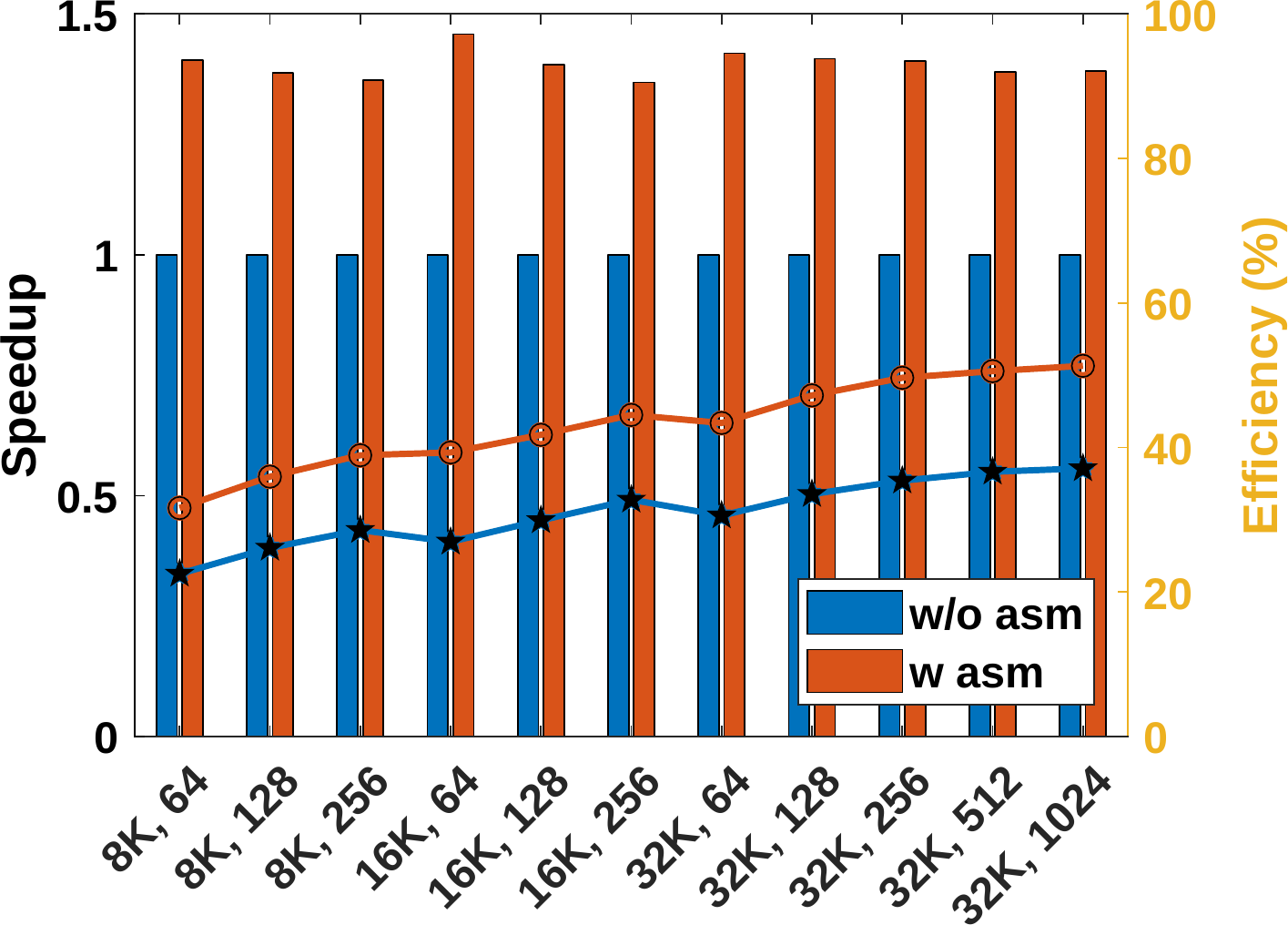}
}
\hspace{-3mm}
\vspace{-3mm}
\subfigure[NTT with explicit dual tile]
{
\includegraphics[width=0.215\textwidth]{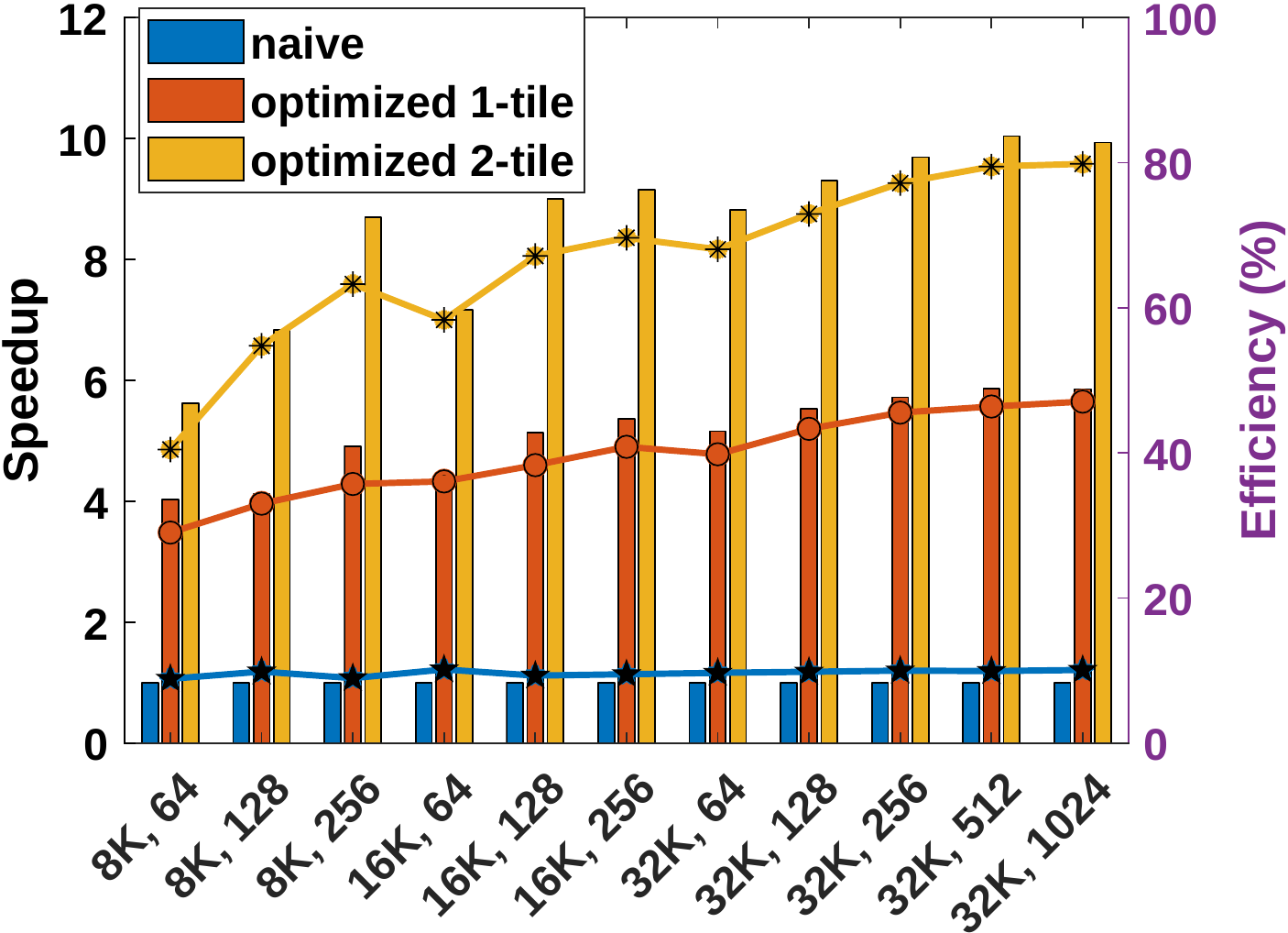}
}
\vspace{-1mm}
\caption{NTT with inline-asm and multi-tile on Device1}
\label{fig:ntt-inline-dual-tile}
\end{figure}

We further introduce assembly-level optimizations to improve the speed of the int64 \textit{add\_mod} and \textit{mul\_mod} ops. As shown in Figure \ref{fig:ntt-inline-dual-tile} (a), these low-level optimizations improve the NTT performance by 35.8\% - 40.7\%, increasing the efficiency of our radix-8 SLM NTT to 47.1\%. The inline assembly low-level optimization provides a relatively stable acceleration percentage for different NTT sizes and instance numbers. This is because assembly-level optimization directly improves the clock cycle of the each 64-bit integer multiplication and addition operation, which is independent of the number of active EUs at runtime.

\subsubsection{Explicit dual-tile submission through DPC++}

With the low-level optimization, we observe that our NTT saturates only up to $47.1$\%, less than half of the peak performance. We observe this low efficiency because DPC++ runtime does not implicitly support multi-tile execution such that only half of the machine has been utilized . To address this issue, we explicitly submit workloads through multiple queues to enable a full utilization of our multi-tile GPU. In regards to the dual-tile peak performance, we manage to reach $79.8$\% of the peak performance through explicit dual-tile submission. Meanwhile, our most optimized NTT is $9.93$-fold faster than the naive baseline for the $32$K-point, $1024$-instance batched NTT.

\subsection{Roofline analysis for NTT}

% New table follows guide at https://people.inf.ethz.ch/markusp/teaching/guides/guide-tables.pdf
\begin{table}[ht] \centering
\caption{Number of 64-bit integer ALU operations of each work-item per round for the NTT}
\begin{tabular}{l
    S[table-format=3] % Align column by 3-digit int
    S[table-format=3] % Align column by 3-digit int
    S[table-format=4] % Align column by 4-digit int
    }
\toprule
\multicolumn{1}{l}{} & \multicolumn{3}{c}{64-bit int ops / round} \\ 
                     & {other}       & {butterfly}       & {total} \\ 
\midrule
radix-2              & 20          & 28              & 48          \\ 
radix-4              & 45          & 112             & 157         \\ 
radix-8              & 120         & 336             & 456         \\ 
radix-16             & 260         & 896             & 1156        \\ 
\bottomrule
\end{tabular}
\label{tab:NTT-op-density}
\end{table}

% Original table
% \begin{table}[ht] \centering
% \begin{tabular}{|c|c|c|c|}
% \hline
% \multicolumn{1}{|l|}{} & \multicolumn{3}{l|}{64-bit int ops / round} \\ \hline
%                       & other       & butterfly       & total       \\ \hline
% radix-2                & 20          & 28              & 48          \\ \hline
% radix-4                & 45          & 112             & 157         \\ \hline
% radix-8                & 120         & 336             & 456         \\ \hline
% radix-16               & 260         & 896             & 1156        \\ \hline
% \end{tabular}
% \caption{Number of 64-bit integer ALU operations of each work-item per round for NTT}
% \label{tab:NTT-op-density}
% \end{table}

\begin{figure}[ht]
\centering
\includegraphics[width=0.40\textwidth]{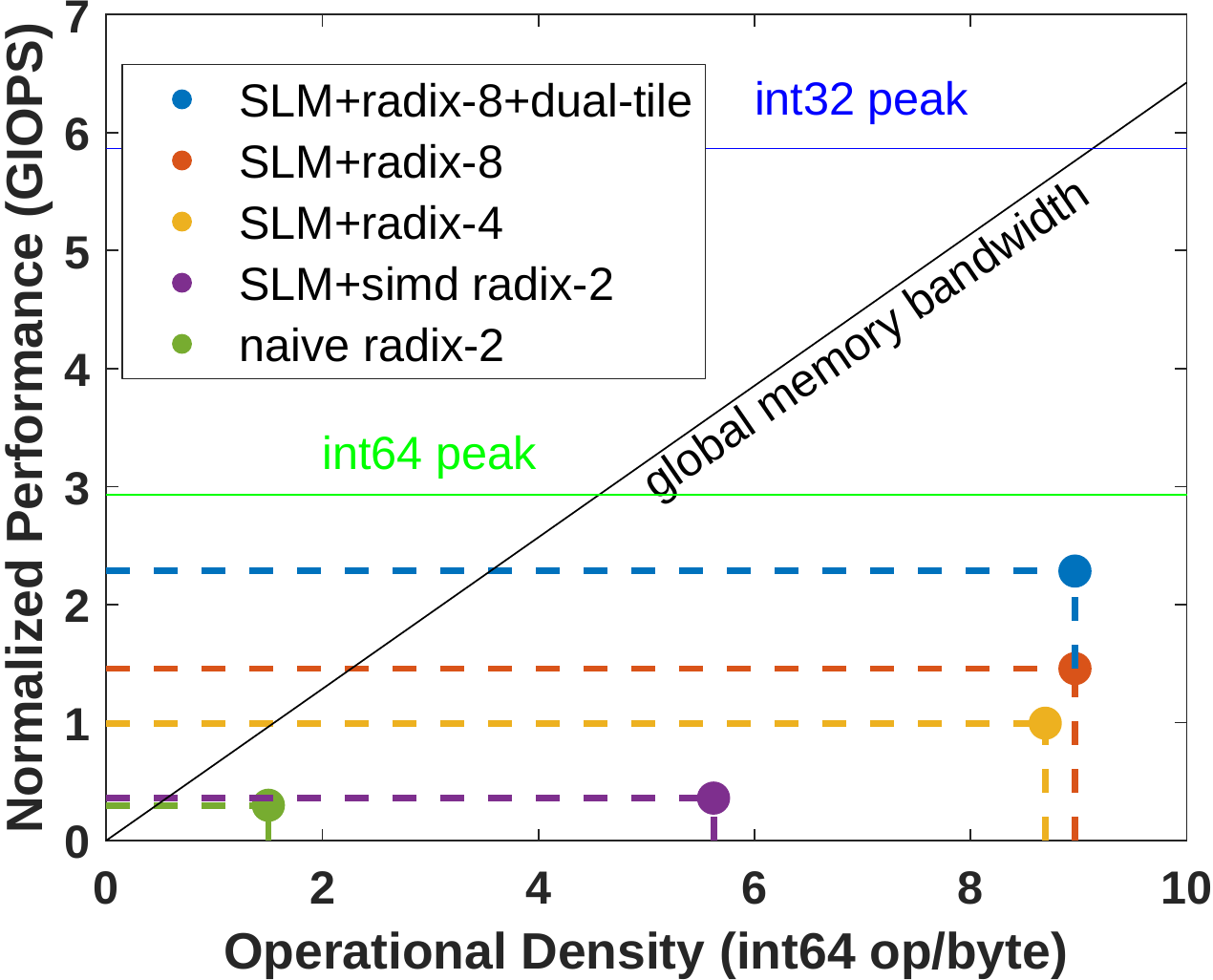}
\caption{Roofline Analysis on Device1}
\label{fig:ntt-roofline}
\end{figure}

The most naive NTT needs to access the global memory for \textit{each} round of the NTT computation. Therefore, its total memory access number can be computed as $2N \log_2(N)$. Here we multiply it by $2$ because of both load and store operations at each round of NTT. Also, we do not count the memory access of last round NTT processing to simplify the analysis and because it is negligible. Table \ref{tab:NTT-op-density} summarizes the number of ALU operations for each NTT variant. \textit{Butterfly} refers to the ALU operations for the NTT butterfly computations while \textit{other} denotes other necessary ALU operations such as index and address pointer computations. The radix-2 NTT performs 48 integer operations for each work-item in a single round of NTT, indicating that the naive NTT consumes $N / 2 \cdot 48 \cdot \log_2(N)$ ALU operations throughout the whole computation process. Further dividing the total ALU number over the total memory access number, one can find that the operational density of naive NTT is equal to $1.5$ for 64-bit integer NTT. This low operational density, as plotted in Figure \ref{fig:ntt-roofline}, suggests that the naive NTT implementation is bounded by the global memory bandwidth and can never reach the int64 peak performance.

On the other hand, for high-radix NTT such as radix-8 staged NTT implementation, we need only two rounds of global memory access for an instance of $32$K-point NTT computation. We first perform one round of radix-8 NTT to reduce the data exchanging gap size from $16$K to $2$K. Right now each work-group computes a $2$K$\times 2$=$4$K-point NTT with all NTT sequence elements held in shared local memory. Therefore, the total global memory access number becomes $2$(LD/ST)$ \times 2$(rounds)$\times$N=$4$N. Considering its total ALU operation number equal to $456$(/round)$\times \log_8(N)$(rounds), one can compute that the operational density of shared local memory radix-8 NTT equals $8.9$, pushing the overall performance to the limits of int64 ALU throughput on Device1. The operational density of other NTT variants are computed similarly.

In addition, a sound operational density with respect to the global memory access does not guarantee satisfactory overall performance. Although the SLM+SIMD radix-2 NTT is no longer bounded by the global memory bandwidth, its practical efficiency remains far from the green line --- int64 peak performance. %Since the radix-16 implementation leads to the register spilling issue, we do not present it in the roofline plot. 

According to the Figure \ref{fig:ntt-roofline}, we conclude that radix-8 shared-local memory NTT with last round kernel fusion enables a sufficient operational density, which allows the performance to be shifted from memory bound to compute-bound. Additionally, the shared local memory utilization together with the low-level optimization for int64 multiplication and DPC++ multi-tile submission, pushes the performance of radix-8 NTT to the ceiling of int64 ALU throughput on Device1.

\subsection{Benchmarking for CKKS HE evaluation routines}

\begin{figure}[ht]
\centering
\includegraphics[width=0.46\textwidth]{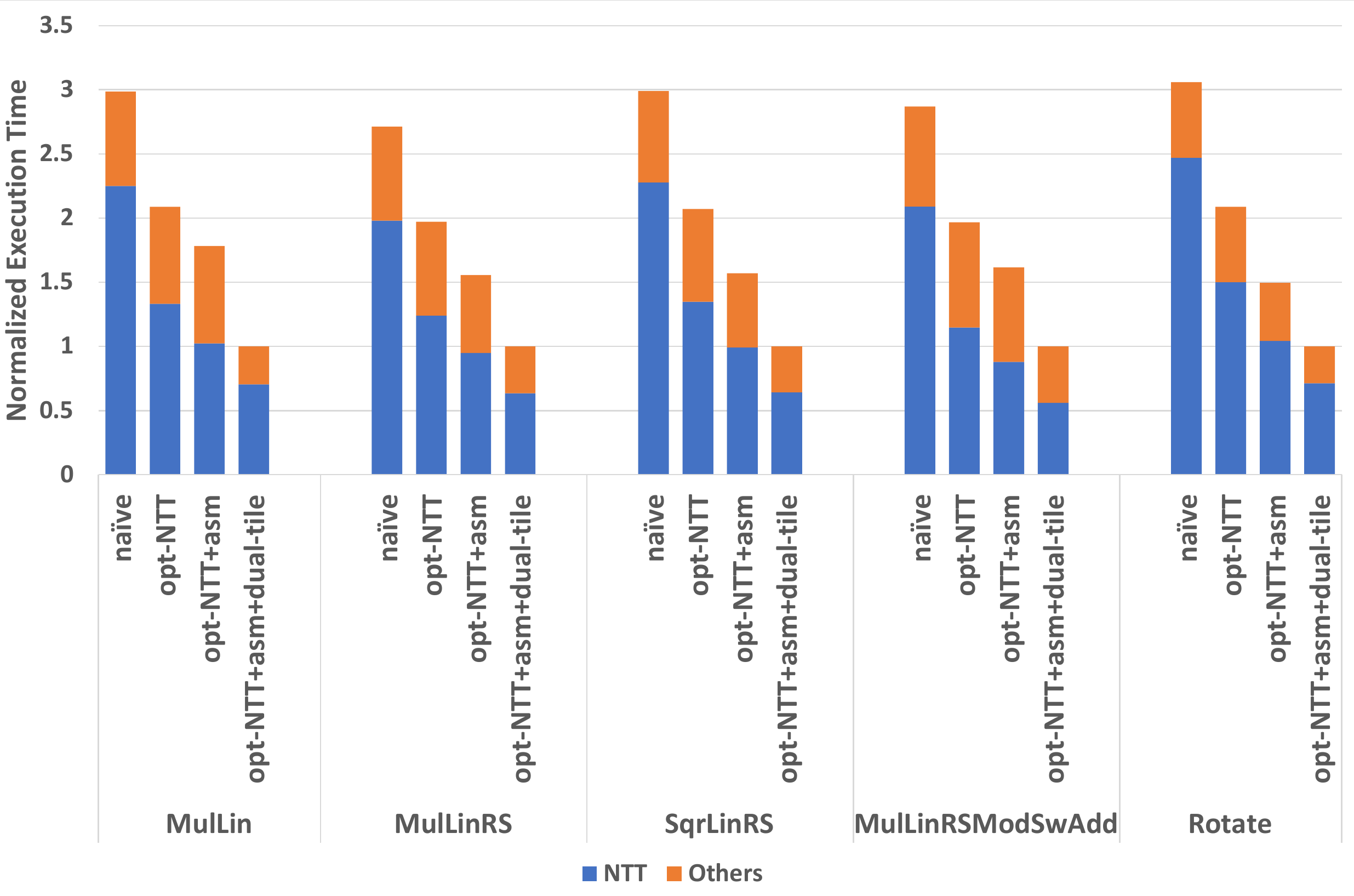}
\caption{Benchmarking HE evaluation routines on Device1}
\label{fig:routine-device1-bench}
\end{figure}

Figure \ref{fig:routine-device1-bench} benchmarks the performance of five basic HE evaluation routines under the CKKS scheme on Device1. Here \textbf{MulLin} denotes a multiplication followed by a relinearization; \textbf{MulLinRS} denotes a multiplication followed by relinearization and rescaling. Relinearization, as we have introduced before, decreases the length of a ciphertext back to 2 after a multiplication. Rescaling is a necessary step for multiplication operation with the goal to keep the scale constant, and also reduce the noise present in the ciphertext. In addition, \textbf{SqrLinRS} refers to a ciphertext square computation with relinearization and rescaling followed. \textbf{MulLinRSModSwAdd} computes a ciphertext multiplication, then relinearizes and rescales it. After this, we switch the ciphertext modulus from $(q_1,q_2,...,q_L)$ down to $(q_1,q_2,...,q_{L-1})$ and scale down the message accordingly. Finally this scaled messaged is added with another ciphertext. The last benchmarked routine, \textbf{Rotate}, rotates a plaintext vector cyclically. We count the GPU kernel time exclusively for routine-level benchmarks. All evaluated ciphertexts are represented as tuples of vectors in $\mathbb{Z}_{q_L}^N$ where $N=32$K and the RNS size is $L=8$.

We present the impact of NTT optimizations to HE evaluation routines in four steps. We first substitute the naive NTT with our radix-8 NTT with SLM. We then employ assembly-level optimizations to accelerate the clock cycle of int64 \textit{add\_mod} and \textit{mul\_mod}. Finally we enable implicit dual-tile submission through the OpenCL backend of DPC++ to fully utilize our wide GPU. The baseline is the naive GPU implementation where no presented optimizations are adopted for the comparison purpose.

The radix-8 NTT with data communications through SLM improves the routine performance by 43.5\% in average. It is worth mentioning that we do not benchmark batched routines and our wide GPU is not fully utilized such that the NTT acceleration is not as dramatic as the results in previous sections. The inline assembly optimization provides a further average 27.4\% improvement in compared with the previous step. Meanwhile, the non-NTT computations show less sensitivity to the inline assembly optimization than the NTT because their computations are typically not as compute-intensive as NTT. We finally submit the kernels through multiple queues to enable the full utilization of our multi-tile GPUs, further improving the performance by 49.5\% to 78.2\% from the previous step, up to 3.05X faster than the baseline.

\subsection{Benchmarking on Device2}

We benchmark our optimizations on another GPU - Device2 which is a single-tile GPU consisting of fewer EUs than Device1. Similar to Device1, the naive radix-2 NTT starts at a ${\sim}15\%$ efficiency of the peak performance, while the shared local memory SIMD implementation either fails to provide significant improvement, but reach only 20.95\%-24.21\% efficiencies. After adopting the radix-8 shared local memory implementation, we manage to obtain a up to 66.8\% of the peak performance, where we are up to 5.47X faster than the baseline at this step. Since Device1 is a single-tile GPU, our final optimization here is to introduce inline assembly to optimize int64 \verb|mul|. Further improving the performance by 28.48\% in average from the previous step, we reach an up to 85.75\% of the peak performance, 7.02X faster than the baseline for 32K-point, 1024-instance NTT.

\begin{figure}[ht]
\centering
\includegraphics[width=0.40\textwidth]{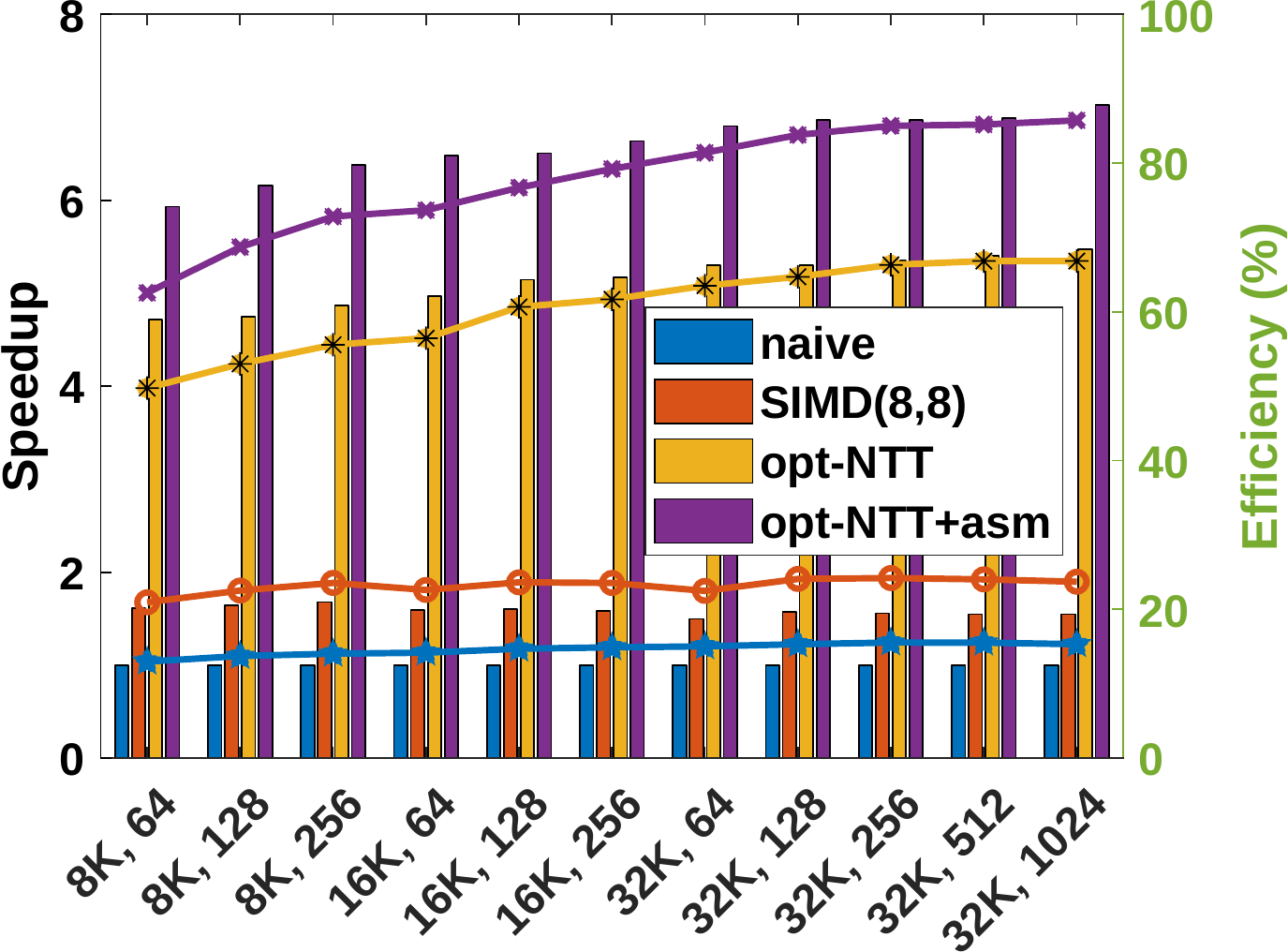}
\caption{Benchmark for NTT on Device2}
\label{fig:ntt-device2-bench}
\end{figure}

\begin{figure}[ht]
\centering
\includegraphics[width=0.46\textwidth]{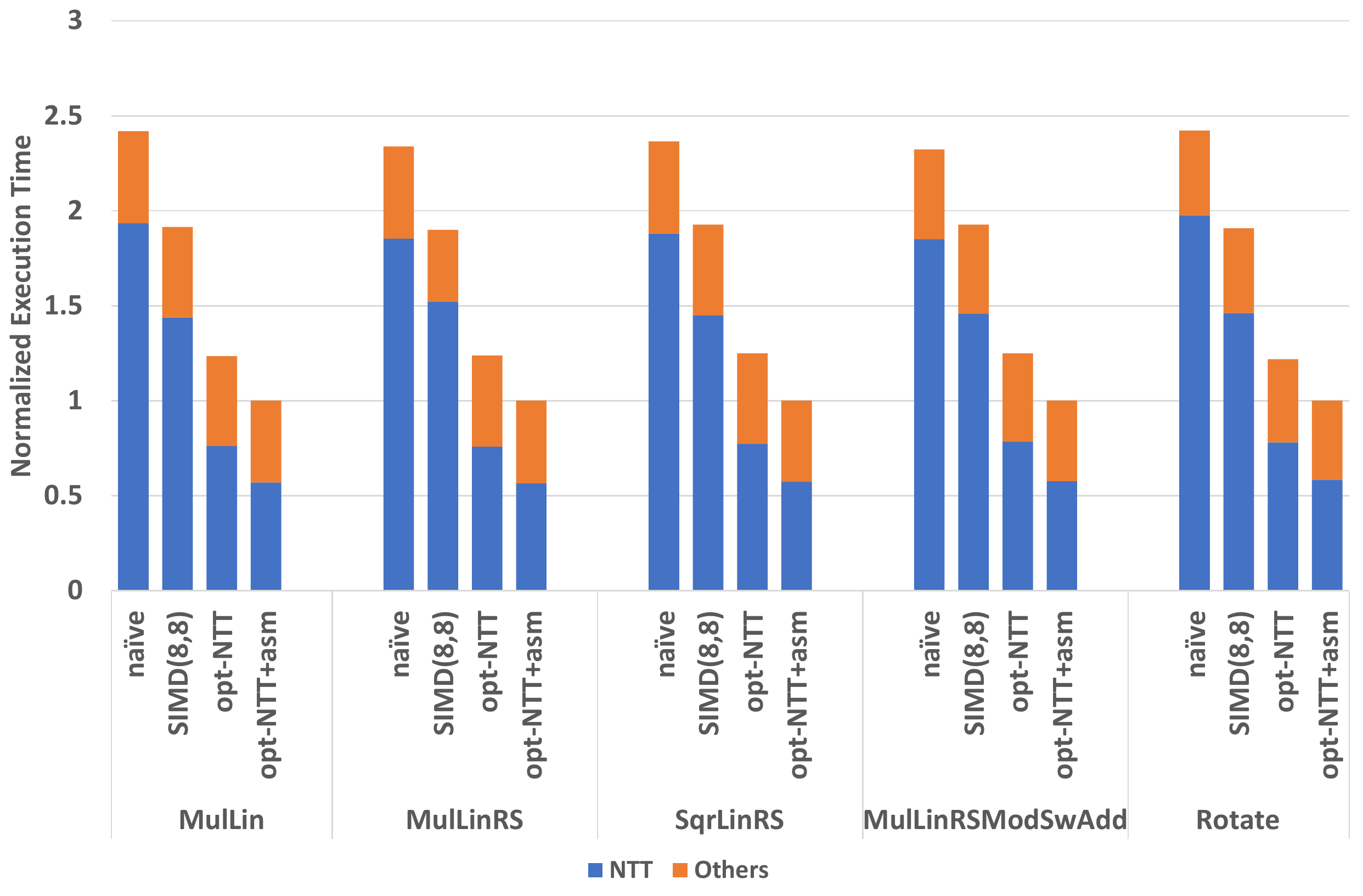}
\caption{Benchmarking HE evaluation routines on Device2}
\label{fig:routine-device2-bench}
\end{figure}

Figure \ref{fig:routine-device2-bench} benchmarks the normalized execution time of HE evaluation routines on Device2. SIMD(8,8) denotes the radix-2 NTT with data exchanging through SLM and SIMD shuffling, where each work-item holds one slot of NTT elements in registers. opt-NTT refers to the optimal NTT variant, radix-8 NTT with data exchanging through SLM, shown in Figure \ref{fig:ntt-device2-bench}. The last step is to further employ inline assembly to optimize modular addition and modular multiplication from instruction level. When substituting the naive NTT using SIMD(8,8) NTT, we observe the execution time of the NTT part is improved by 34\% in average while the overall routines are accelerated by 29.6\%. When switching to our optimal NTT variant, we observe the overall performance becomes faster than the baseline by 1.92X in average. Further enabling assembly-level optimizations, we manage to reach 2.32X - 2.41X acceleration for all five HE evaluation routines on this single-tile Intel GPU.

\subsection{Benchmarks for polynomial matrix multiplication}

\begin{figure}[ht] \centering
\vspace{-2mm}
% \hspace{-1mm}
\subfigure[{Device 1}]
{
\includegraphics[width=0.215\textwidth]{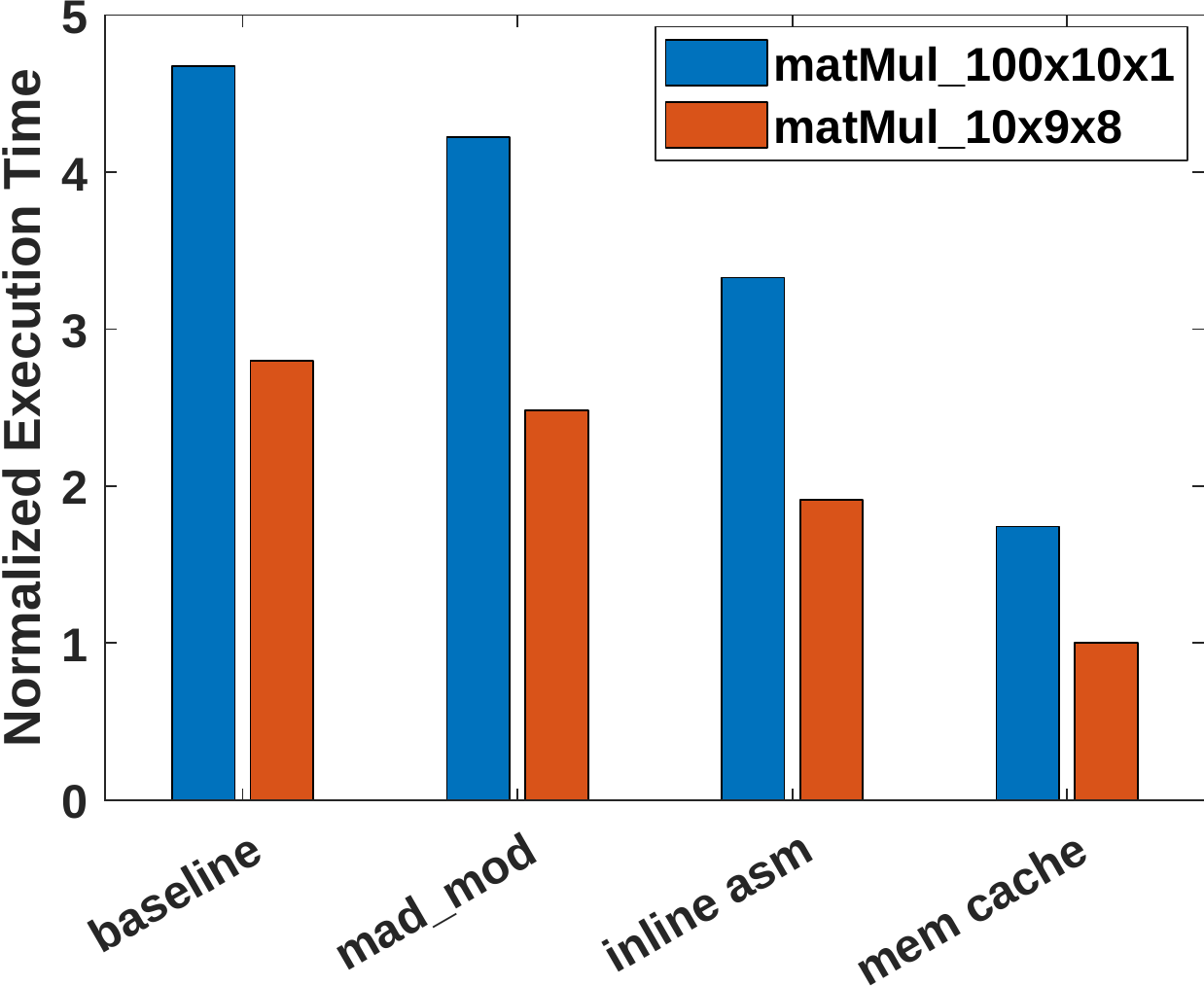}
}
\hspace{-3mm}
\vspace{-3mm}
\subfigure[Device 2]
{
\includegraphics[width=0.215\textwidth]{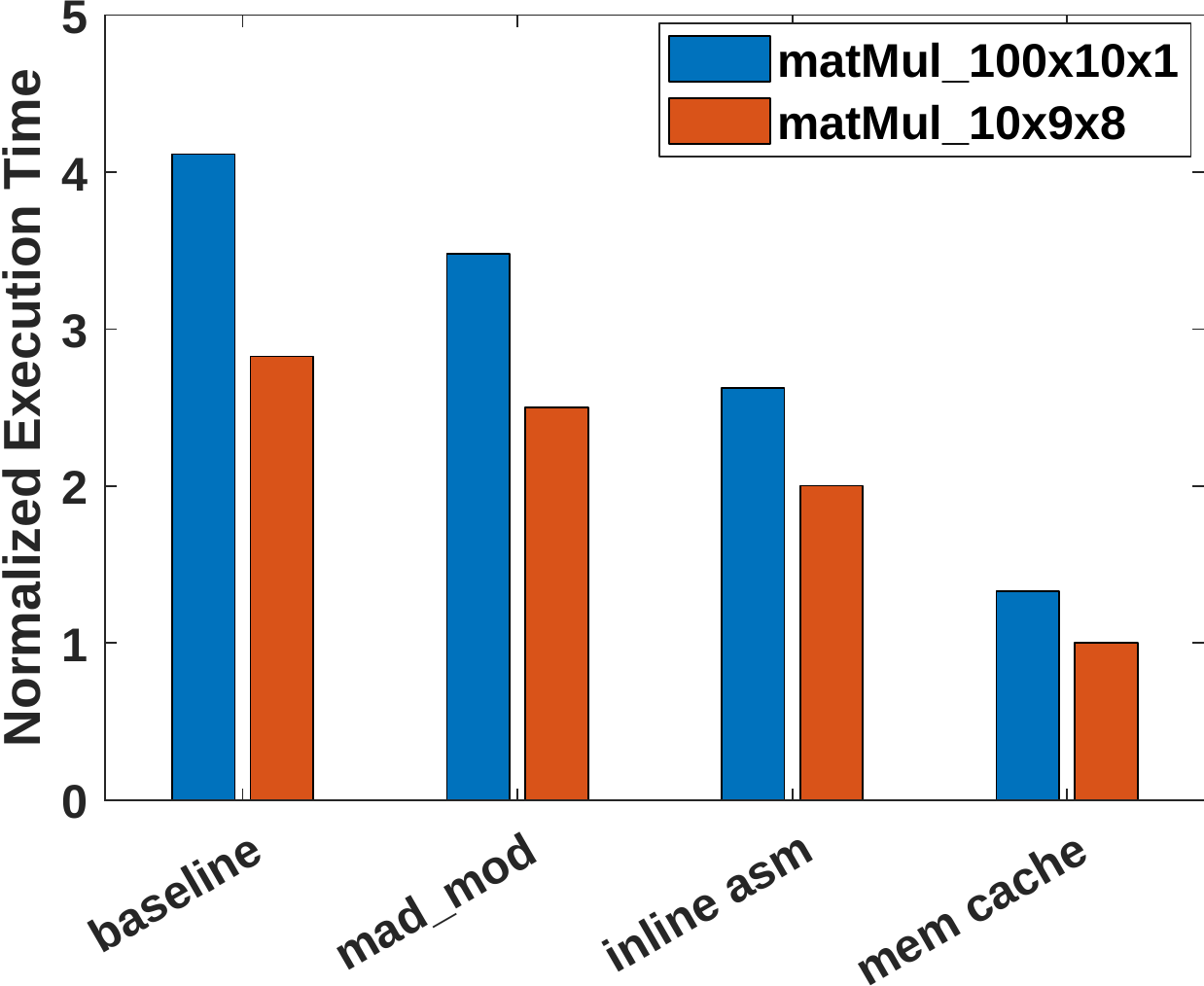}
}
\vspace{-1mm}
\caption{Element-wise polynomial multiplication}
\label{fig:matmul}
\end{figure}

Besides the algorithmic-level optimizations, we also demonstrate our instruction-level and application-level optimizations, which are modulus fusion, inline assembly for int64 multiplication and memory cache using a representative application of HE, encrypted element-wise polynomial matrix multiplication. In Figure \ref{fig:matmul}, \verb|matMul_mxnxk| denotes a matrix multiplication $C$+=$A*B$, where $C$ is $m$-by-$n$, A is $m$-by-$k$ and B is $k$-by-$n$. Each matrix element is an 8K-element plaintext polynomial so each \textit{element-wise} multiplication of \verb|matMul| is a polynomial multiplication. Modulo operations are always applied at the end of each multiply or addition between polynomial elements. Before starting \verb|matMul|, we need to allocate memory, initialize, encode and encrypt input sequences. Once \verb|matMul| is completed, we decrypt the computing results and count the elapsed time of the whole process.

Figure \ref{fig:matmul} compares our instruction-level and application-level optimizations for \verb|matMul| on both Device1 and Device2. The fused \verb|MAD_MOD| and inline assembly accelerate the both 100x100x1 and 10x9x8 polynomial matrix multiplications by 11.8\% and 28.2\%, respectively, in average on Device1. With the memory cache introduced, both \verb|matMul| applications are further improved by ${\sim}90\%$. Our systematic optimizations accelerate two \verb|matMul| benchmarks 2.68X and 2.79X, respectively, compared with the baseline on Device1. In regards to Device2, we observe a similar trend. These three optimizations together provide us with 3.11X and 2.82X acceleration for two \verb|matMul| tests over the baseline on this smaller GPU.

\section{Conclusions} \label{sec:conclusion}

In this paper, we design and develop the first-ever SYCL-based GPU backend for Microsoft SEAL APIs. We accelerate our HE library for Intel GPUs spanning assembly level, algorithmic level and application level optimizations. Our optimized NTT is faster than the naive GPU implementation by 9.93X, reaching up to 85.1\% of the peak performance. In addition, we obtain up to 3.1X accelerations for HE evaluation routines and the element-wise polynomial matrix multiplication application. Future work will focus on extending our HE library to multi-GPU and heterogeneous platforms.

\bibliographystyle{./bibliography/IEEEtran}
\bibliography{./bibliography/IEEEexample}

\end{document}